\documentclass[reprint,amsfonts, amssymb, amsmath,  showkeys, nofootinbib,pra, superscriptaddress, twocolumn,longbibliography]{revtex4-2}

\makeatletter
\let\newfloat\newfloat@ltx
\makeatother
\usepackage[english]{babel}
\usepackage[normalem]{ulem}
\usepackage[utf8]{inputenc}
\usepackage{graphics}
\usepackage{selinput}

\usepackage{braket}
\usepackage{amsthm}
\usepackage{mathtools}
\usepackage{physics}
\usepackage[dvipsnames]{xcolor}
\usepackage{graphicx}
\usepackage[left=16mm,right=16mm,top=35mm,columnsep=15pt]{geometry} 
\usepackage{adjustbox}
\usepackage{placeins}
\usepackage[T1]{fontenc}
\usepackage{lipsum}
\usepackage{csquotes}
\usepackage{mathbbol} 
\usepackage{notes2bib}
\usepackage{multirow}
\usepackage[linesnumbered,ruled,vlined]{algorithm2e}
\SetKwInput{kwInit}{Init}

\def\LC{\mathcal{L}}

\usepackage[makeroom]{cancel}
\usepackage[toc,page]{appendix}
\usepackage[colorlinks=true,citecolor=blue,linkcolor=magenta]{hyperref}

\usepackage{tikz}
\tikzset{every picture/.style=remember picture}

\DeclarePairedDelimiterX{\infdivx}[2]{(}{)}{%
  #1\;\delimsize\|\;#2%
}

\newcommand{\Ubb}{\mathbb{U}}

\newcommand{\AC}{\mathcal{A}}
\newcommand{\BC}{\mathcal{B}}

\newcommand{\FC}{\mathcal{F}}

\newcommand{\OC}{\mathcal{O}}
\newcommand{\PC}{\mathcal{P}}

\newcommand{\Var}{{\rm Var}}

\renewcommand{\geq}{\geqslant}
\renewcommand{\leq}{\leqslant}

\DeclareMathOperator*{\argmin}{arg\,min}
\renewcommand{\vec}[1]{\boldsymbol{#1}}  

\newcommand{\bs}{\textsf{BS}}

\newcommand{\thv}{\vec{\theta}}

\def\be{\begin{equation}}
\def\ee{\end{equation}}
\def\bs{\begin{split}}
\def\e{\end{split}}
\def\ba{\begin{eqnarray}}
\def\bea{\begin{eqnarray}}

\def\tea{\end{eqnarray}}
\def\ea{\end{eqnarray}}
\def\eea{\end{eqnarray}}

\newtheorem*{proposition*}{Proposition}

\usepackage{amssymb}
\usepackage{dsfont}

\usepackage[normalem]{ulem}
\usepackage{hyperref}

\def\be{\begin{equation}}
\def\te{\end{equation}}
\def\ee{\end{equation}}
\def\ba{\begin{eqnarray}}
\def\bea{\begin{eqnarray}}

\def\tea{\end{eqnarray}}
\def\ea{\end{eqnarray}}
\def\eea{\end{eqnarray}}

\begin{document}

\title{Connecting phases of matter to the flatness of the loss landscape in \\  analog variational quantum algorithms }

\author{Kasidit Srimahajariyapong}
\email{kasidit.quantum@gmail.com}
\affiliation{Chula Intelligent and Complex Systems Lab, Department of Physics, Faculty of Science, Chulalongkorn University, Bangkok, Thailand, 10330}
\affiliation{Siam Quantum Square, Faculty of Science, Chulalongkorn University, Bangkok, Thailand, 10330}

\author{Supanut Thanasilp}
\email[Correspondence to: ]{supanut.thanasilp@gmail.com}
\affiliation{Chula Intelligent and Complex Systems Lab, Department of Physics, Faculty of Science, Chulalongkorn University, Bangkok, Thailand, 10330}
\affiliation{Siam Quantum Square, Faculty of Science, Chulalongkorn University, Bangkok, Thailand, 10330}
\affiliation{Centre for Quantum Science and Engineering, Ecole Polytechnique Fédérale de Lausanne (EPFL), CH-1015 Lausanne, Switzerland}

\author{Thiparat Chotibut}
\email[Correspondence to: ]{thiparatc@gmail.com}
\affiliation{Chula Intelligent and Complex Systems Lab, Department of Physics, Faculty of Science, Chulalongkorn University, Bangkok, Thailand, 10330}
\affiliation{Siam Quantum Square, Faculty of Science, Chulalongkorn University, Bangkok, Thailand, 10330}

\date{\today}

\begin{abstract}
Variational quantum algorithms (VQAs) promise near‑term quantum advantage, yet parametrized quantum states commonly built from the digital gate-based approach often suffer from scalability issues such as barren plateaus, where the loss landscape becomes flat.  We study an \emph{analog} VQA ansätze composed of $M$ quenches of a disordered Ising chain, whose dynamics is native to several quantum simulation platforms.  By tuning the disorder strength we place each quench in either a thermalized phase or a many‑body‑localized (MBL) phase and analyse (i) the ansätze's expressivity and (ii) the scaling of loss variance.  Numerics shows that both phases reach maximal expressivity at large $M$, but barren plateaus emerge at far smaller $M$ in the thermalized phase than in the MBL phase.  Exploiting this gap, we propose an MBL initialisation strategy: initialise the ansätze in the MBL regime at intermediate quench $M$, enabling initial trainability while retaining sufficient expressivity for subsequent optimization.  The results link quantum phases of matter and VQA trainability, and provide practical guidelines for scaling analog‑hardware VQAs.

\end{abstract}

\maketitle

\section{Introduction}
Variational quantum algorithms (VQAs) offer a promising near-term approach to quantum advantage and have already addressed modest-scale optimization tasks in quantum chemistry \cite{peruzzo2014variational,google2020hartree,mcardle2020quantum}, combinatorial optimization \cite{farhi2014quantum,abbas2024challenges,tangpanitanon2023hybrid}, high-energy physics \cite{di2024quantum}, and quantum machine learning \cite{schuld2015introduction,biamonte2017quantum,benedetti2019parameterized,cerezo2022challenges}. This hybrid approach combines a quantum processor to prepare a parametrized quantum state with a classical optimizer that adjusts those parameters to minimize a problem-specific loss \cite{cerezo2020variationalreview,bharti2021noisy}. Yet scaling VQAs to large system sizes where quantum advantage could arise is far from trivial. Barriers to scalability include proliferation of spurious local minima \cite{anschuetz2021critical, anschuetz2022quantum}, efficient classical surrogatability \cite{schreiber2022classical,sweke2023potential,sahebi2025dequantization}, and most prominently the \emph{barren plateau} (BP) phenomenon: the loss landscape becomes exponentially flat (together with vanishing loss gradients) in the system size, reflecting the curse of dimensionality in the space where the parametrized states operate~\cite{mcclean2018barren,larocca2024review,cerezo2025does}. In a BP regime one needs exponentially many measurement shots to reliably navigate through the flat landscape regions, rendering optimizing VQAs impractical at scale~\cite{arrasmith2020effect}.

While these parametrized states are typically built through a sequence of quantum gates (digital, gate-based VQAs) and can achieve universality with deep circuits, the deep circuits' \emph{expressivity} (i.e., the ability to explore vast regions of Hilbert space) can be excessive for a given task and has been identified as a root cause of BPs~\cite{mcclean2018barren,fontana2023theadjoint,ragone2023unified,larocca2024review}. A complementary approach is to utilize \emph{native} many‑body
dynamics of \emph{analog} quantum simulators as a quantum processor, letting the hardware evolve under its natural, yet controllable, quantum evolution. Indeed, state‑of‑the‑art experimental platforms such as Rydberg arrays, trapped ions, and superconducting circuits can realize target Hamiltonian with high fidelity and provide tunable controls that help probing quantum phases of matter \cite{bernien2017probing,zhang2017observation,gross2017quantum,elben2020cross,semeghini2021probing,braumuller2022probing,shaw2024benchmarking}. A striking demonstration is the observation of many‑body localization (MBL) phase in a $125$‑site two‑dimensional lattice \cite{choi2016exploring}. Whereas thermalized systems equilibrate \cite{rigol2008thermalization,gogolin2016equilibration,d2016quantum,mori2018thermalization}, MBL states retain local memory due to strong disorder and interactions and are expected to survive in the thermodynamic limit \cite{vznidarivc2008many,pal2010many,serbyn2013local,serbyn2013universal,huse2014phenomenology,serbyn2014quantum,abanin2019colloquium}. MBL signatures have since been reported across multiple hardware platforms \cite{schreiber2015observation,bordia2016coupling,smith2016many,roushan2017spectroscopic,xu2018emulating,lukin2019probing}, showcasing the ability of analog devices to simulate complex, classically intractable dynamics while remaining experimentally controllable.

Motivated by these experimental advances, a growing body of work seeks to merge analog dynamics with the variational framework. Pulse‑level and quench‑based ans\"{a}tzes can be optimized \emph{in situ}, turning the tunable simulator itself into a variational quantum processor and bypassing gate-decomposition overheads~\cite{khaneja2005optimal,tangpanitanon2020expressibility,thanasilp2021quantum,meitei2021gate,magann2021pulses,michel2023blueprint,liang2023hybrid,asthana2023leakage,de2023pulse,egger2023pulse,meirom2023pansatz,liang2024napa}. Such schemes align with broader digital-analog approaches~\cite{parra2020digital,garcia2024digital} and exploit hardware‑native interactions that naturally realize non‑trivial quantum phases~\cite{khaneja2005optimal,magann2021pulses,meitei2021gate,de2023pulse}. Initial studies show that the underlying phase can significantly influence both expressivity and trainability~\cite{tangpanitanon2020expressibility}, although how these effects scale with system size remains largely unexplored. In addition, in the gate-based setting, MBL-inspired initialisations have also been proposed to mitigate BPs~\cite{park2024hardware,cao2024exploiting,xin2024improve}. These observations raise two key questions: \emph{Which phases of matter are favourable for analog VQAs, and how can they be
leveraged to mitigate BPs?}

\begin{figure*}[htbp]
    \centering
    \includegraphics[width=0.9\textwidth]{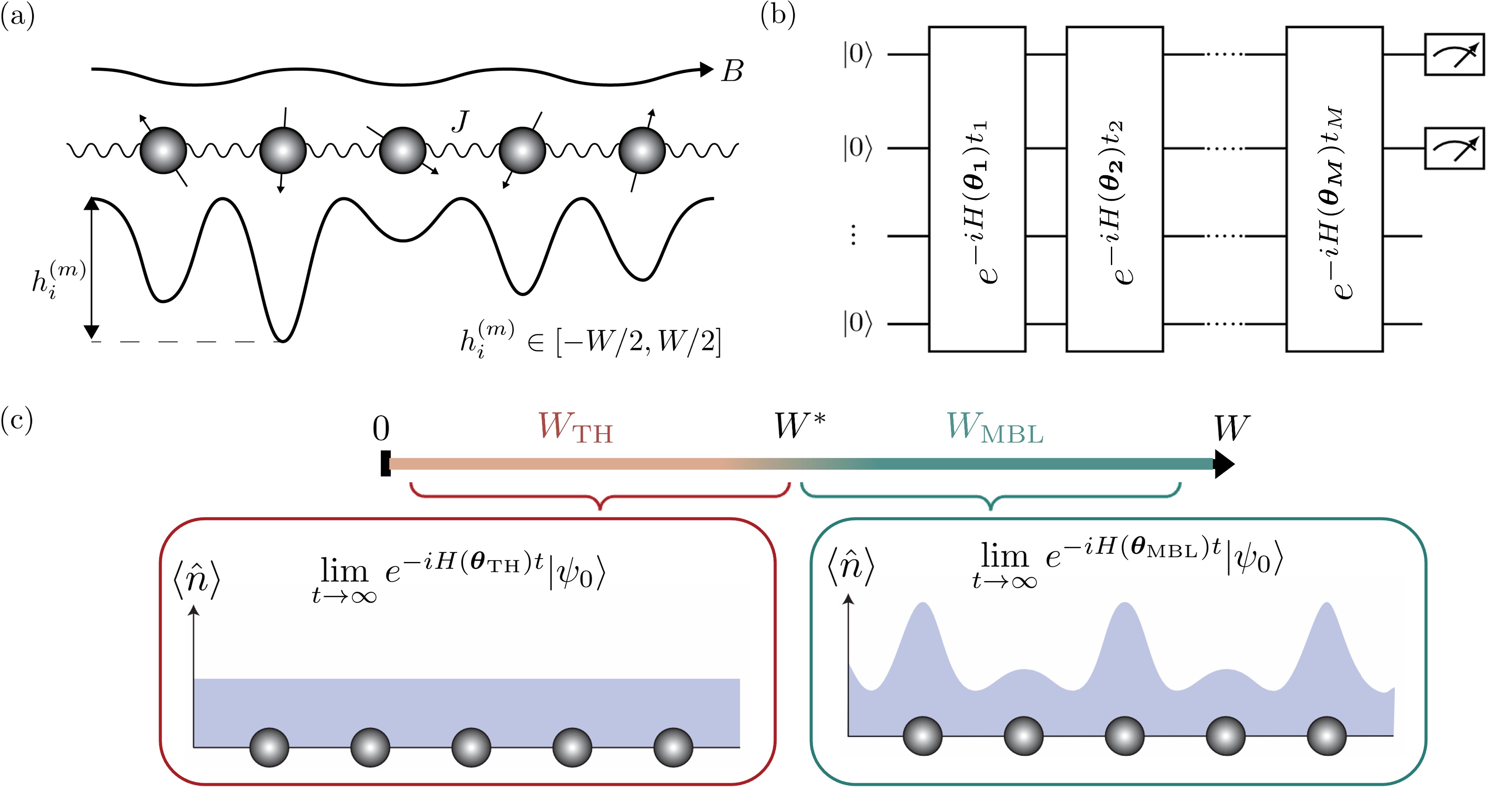}
     \caption{
     \textbf{Overview of our analog VQA framework} (a) Schematic representation of a periodic Ising spin chain with nearest-neighbour interactions $J$, disordered on-site energy $h_i^{(m)}$ at the $m$-th quench, and a transverse field $B$. Each on-site disorder energy is uniformly drawn from the interval $[-W/2,W/2],$ with $W$ controlling the disorder strength. 
     (b) Analog VQA modeled as a series of quench dynamics. Each block represents a unitary evolution under the quench Hamiltonian $H(\vec{\theta}_m)$ for a time $t_m$, parametrized by the disorder configuration $\vec{\theta}_m = \{h_i^{(m)}\}$ and measured by local observables. While the evolution times can in general be trainable parameters, we fix them here for our investigation on the connection with phases but they can be made trainable when solving the actual optimization problems (see the Appendix).  
     (c) A schematic phase diagram of the system as a function of disorder strength \( W \) (defined for a sufficiently long evolution time of a single quench dynamics) shows a transition between the many-body localized (MBL) phase and the thermalized phase. Bottom panels depict spin density profiles after long-time evolution: in the MBL phase (\( W < W^* \)), the system retains memory of the initial state, while in the thermalized phase (\( W > W^* \)), the system relaxes to a thermal state. This setup enables us to study the  interplay between disorder-induced quantum phases and the performance and scalability of analog VQAs setup in (b).}
    \label{fig:set-up}
\end{figure*}

In this work, we tackle these questions using an analog ansätze built from $M$ global quenches of a disordered nearest‑neighbour Ising chain (Fig.~\ref{fig:set-up}).  By tuning the on‑site disorder each quench can be placed in either a thermalized or an MBL phase, allowing us to probe how phase structure shapes two key VQA properties: expressivity and the scaling of loss variances, and propose practical BP-mitigating initialisation strategy. 

\begin{figure*}[htbp]
    \centering
    \includegraphics[width=0.95\textwidth]{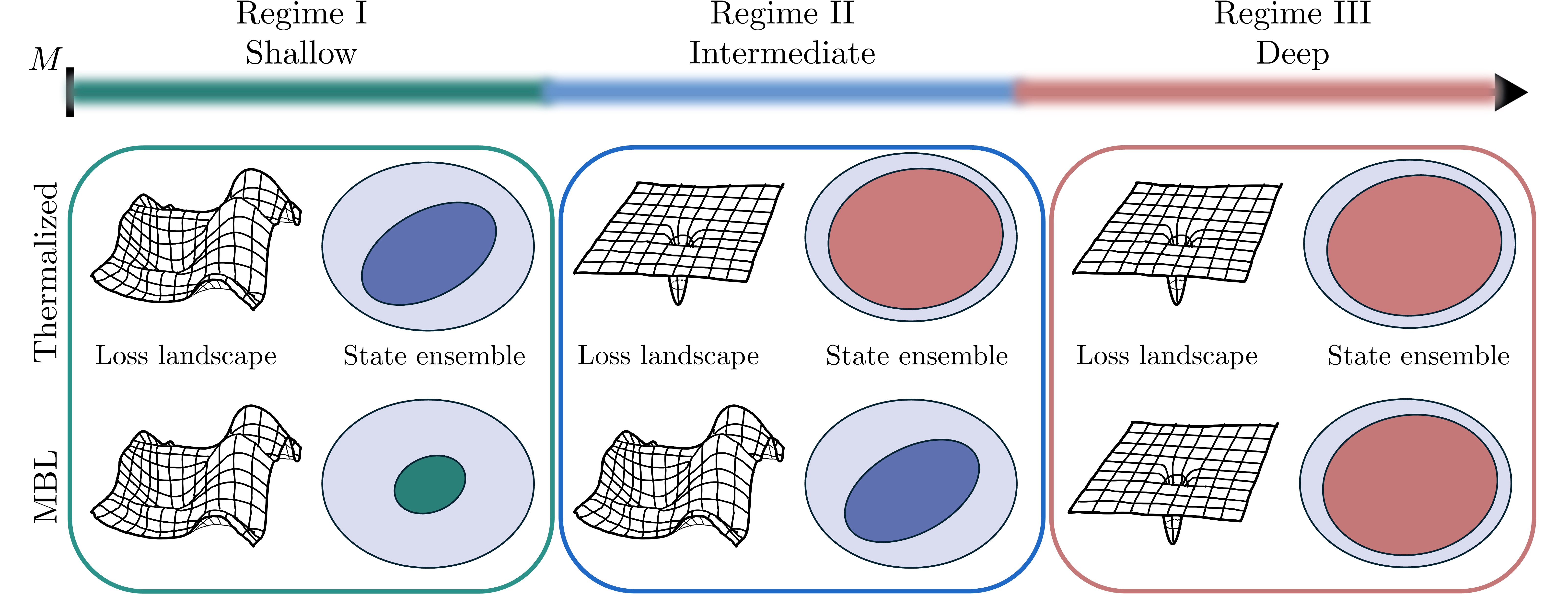}
     \caption{\textbf{Summary of Results.} A visualization of Hilbert space exploration and loss landscapes for our initialisation strategies using two different phases of matter: thermalized and MBL. The results are categorized by the number of quenches (analogous to the circuit depth in gate-based VQA approaches): shallow, intermediate, and deep. \textbf{Shallow quenches} -- Both initialisations exhibit non-flat loss landscapes. Thermalized dynamics lead to faster state evolution in Hilbert space compared to MBL dynamics, but the model expressivity in both initialisations is far from maximal. \textbf{Intermediate quenches} -- the thermalized initialisation begins to exhibit the barren plateaus and achieves maximal expressivity, while the MBL loss landscape remains non-flat and Hilbert space is not yet completely explored. This intermediate quench regime highlights our proposed initialisation strategy: {\it using MBL initialisation strategy at intermediate quenches allows the model to attain high expressivity while retaining trainability}. \textbf{Deep quenches} -- For both initialisations, barren plateaus appear and the Hilbert space is fully explored, indicating maximal expressivity.}
    \label{fig:summary}
\end{figure*}

\paragraph*{How quantum phases shape fundamental properties of analog VQAs.} Our results show that both phases become maximally expressive in the deep‑quench limit, but BPs appear at far fewer quenches $M$ in the thermalized phase than in the MBL phase.  This difference stems from the underlying many-body dynamics; significantly larger portion of the Hilbert space is explored 
during each quench in the thermalized phase. In line with the connection between expressivity and the vanishing loss gradients in Ref.~\cite{holmes2021connecting}, we observe that loss variances shrink as the number of quenches makes the ansätze approach a unitary 2‑design, irrespective of the quantum phase. 
This relationship is summarized in Fig.~\ref{fig:summary}. Further investigation into the entanglement entropy growth in each phase well aligns with the trend observed in the flatness of the loss landscape.

\paragraph*{MBL initialisation strategy.} The separation of expressivity scaling suggests a practical initialisation strategy. Initialise in the MBL ansätze at an \emph{intermediate} number of quenches $M$ such that it is deep enough that the corresponding thermalized phase is maximally expressive, yet shallow enough to initially avoid BPs in the MBL phase. This strategy provides significant gradients at the first optimization step while retaining a sufficient expressive power for subsequent optimization.  We benchmark the strategy on small proof‑of‑principle tasks (ground‑state VQE and a Max‑Cut problem of random-weight complete graphs).  This MBL initialisation strategy can be viewed as an analog counterpart of near‑identity initialisation for gate-based circuits \cite{grant2019initialization,zhang2022escaping}. 

Our approach aims to bridge digital and analog variational framework and highlight the role of quantum phases of matter in variational algorithms.  We expect this approach to be testable on current analog hardware and to inform the design principles of scalable quantum phase‑aware analog VQAs.

\section{Framework}
\label{sec:framework}
\subsection{Analog simulator as a VQA ans\"{a}tze}
\label{sec:VQA}
We commence by introducing a series of quench dynamics as an ans\"{a}tze for analog VQAs, serving as a framework to study how the dynamical properties of different quantum phases of matter can influence analog VQA performance.

Given an initial state $\ket{\psi_0}$ and some parametrized quantum dynamics $U(\vec{\theta})$ on $n$ qubits with trainable parameters $\thv$, 
we consider the loss function which is defined as the expectation value of an observable $O$ 
\begin{equation}
\label{eq:loss_f}    \LC(\vec{\theta}) = \bra{\psi_{0}}U^\dagger(\vec{\theta})OU(\vec{\theta})\ket{\psi_{0}} \;.
\end{equation} 

By using a classical optimizer to navigate through the loss function landscape, the task is to optimize the loss function 
\begin{align}
    \thv^* = \argmin_{\thv} \LC(\thv) \;.
\end{align}

In this work, the parametrized quantum dynamics is chosen to be a series of $M$ quenches where each quench is a quantum evolution under the system's Hamiltonian $H(\thv_m)$ characterized by parameters $\thv_m$, with an evolution time $t_m$. The overall unitary evolution is given by
\begin{equation}
    \label{eq:quench_dynamics}
    U(\vec{\theta}) =  \prod_{m=1}^M  e^{-i H(\thv_m)  t_m} \;\;.
\end{equation}
The model Hamiltonian we study is a disordered Ising spin chain with nearest-neighbour interactions under periodic boundary conditions, given by
\begin{equation}
    \label{eq:IsingModel}
        H(\vec{\theta}_m) =  J\sum_{i=1}^n Z_i Z_{i+1} + B \sum_{i=1}^n X_i + \sum_{i=1}^n h_{i}^{(m)} Z_i \;,
\end{equation} 
where $Z_i$ and $X_i$ are the single-qubit Pauli-Z and Pauli-X operators acting on qubit $i$, respectively, $J$ is the interaction strength, $B$ is the transverse field strength, and $\{ h_{i}^{(m)} \}_{i=1}^n$ are the on-site disorder energies at the $m$-th quench, which are treated as trainable parameters $\thv_m$. The periodic boundary conditions are implemented by identifying $Z_{n+1} = Z_1$ in the interaction term. 
When analysing the fundamental VQA properties, we fix the evolution time for each quench to be $1/J$, but later on for solving actual optimization tasks these times can be treated also as additional trainable parameters. Note that the choice of evolution time is not unique. An alternative timescale of interest is, for example, $n/J$, which should be sufficient to generate correlations between the first and last qubits in this topology within a single quench. We explore this case in Appendix~\ref{appx:diff_time}.

We initialise the spin configuration in the all-zero state $\ket{\psi_0} = \ket{\vec{0}}$ in the computational basis. For each on-site disorder energy, $h_{i}^{(m)}$ will initially be drawn randomly from a uniform distribution over the interval $\left[-\frac{W}{2}, \frac{W}{2}\right]$, where $W$ represents the disorder strength. By adjusting the disorder strength, each quench dynamics can explore different phases of matter, as will be discussed in the next subsection.

This model serves as a simplified version of the trapped ion Hamiltonian used in experiments like those in Ref.~\cite{smith2016many}, but without long-range interactions. It has been employed in previous studies for investigating quantum generative modeling~\cite{tangpanitanon2020expressibility}, phase transitions~\cite{thanasilp2021quantum}, as well as sampling quantum advantage~\cite{tangpanitanon2023signatures}. While our focus is on this simplified model, we expect that the results obtained will generalize to other systems as well. Indeed, in Appendix~\ref{appx:MoreRealisticModel}, we extend our analysis to include the long-range interaction version of the model.

\subsection{Phases of matter}
\label{subsec:phase}

Depending on the disorder strength $W$, the long-time evolution under the Hamiltonian in Eq.~\eqref{eq:IsingModel} (i.e., $\lim\limits_{t \rightarrow \infty} e^{-iH(\thv_0)t}|\psi_0\rangle$ for a given disorder configuration $\thv_0$) can exhibit two distinct phases of matter: (i) the thermalized phase and (ii) the many-body localized (MBL) phase. 

In the weak disorder limit, the system thermalizes under its own dynamics to some finite temperature and is said to obey the Eigenstate Thermalization Hypothesis (ETH) \cite{deutsch1991quantum, srednicki1999approach, rigol2008thermalization, deutsch2018eigenstate}. According to the ETH, the eigenstates of the Hamiltonian behave as thermal states, meaning that any subsystem becomes thermalized due to interactions with the rest of the system. This leads to a volume law for entanglement entropy \cite{eisert2010colloquium}, where the entanglement entropy scales linearly with sub-system's size and the entanglement growth is linear in time. 

Conversely, strong disorder can prevent the system from thermalizing, leading to the MBL phase \cite{alet2018many,abanin2019colloquium}. This non-ergodic behaviour is associated to the emergence of local integrals of motion \cite{serbyn2013local}, resulting in 
uncorrelated localized eigenstates \citep{oganesyan2007localization,pal2010many}. Consequently, the entanglement entropy in the MBL phase grows relatively slow only logarithmically over time, and the dynamics do not erase initial state information~\cite{serbyn2013universal}. Notably, the MBL phase is robust against local perturbations and persists in the thermodynamic limit $n\rightarrow \infty$.

A standard approach to distinguish these two phases involves analysing the level statistics of the Hamiltonian \cite{oganesyan2007localization, atas2013distribution}. Here, one  computes the ratios of consecutive level spacings from the eigenenergies defined as
\begin{equation}
    r_i = \frac{\min(\Delta_i,\Delta_{i+1})}{\max(\Delta_i,\Delta_{i+1})},
\end{equation} 
with $\Delta_i = \epsilon_{i+1} - \epsilon_{i}$ and the eigenenergies $\{ \epsilon_i \}_i$ are ordered such that $\epsilon_{i+1} \geq \epsilon_{i}$. Then, the distribution of these ratios $\{ r_i \}_i$ is referred to as the level statistics ${\rm Pr}(r)$.

In the thermalized phase, the level statistics follows that of the Gaussian Orthogonal Ensemble (GOE):
\begin{align}
    {\rm Pr}_{\rm GOE}(r) = \frac{27}{4}\frac{r+r^2}{(1+r+r^2)^{5/2}},
\end{align}
which characterizes the statistics of eigenvalues of random matrices with the orthogonality constraint. This connection to the random matrix ensemble partially justifies the chaotic behavior of the dynamics in this phase. Additionally, ${\rm Pr}_{\rm GOE}(0) = 0$ indicates level repulsion, implying that eigenvalues avoid crossing and are correlated.

In contrast, the level statistics in the MBL phase follows the Poisson distribution,
\begin{align}
    {\rm Pr}_{\rm POI}(r) = \frac{2}{(1+r)^2}.
\end{align}
The absence of level repulsion at $r=0$ implies that eigenvalues are uncorrelated and can be arbitrarily close to each other, allowing level crossings. This behavior reflects the localized nature of the eigenstates in the MBL phase; localized eigenstates do not overlap significantly, so the energy levels are determined almost independently of one another.

\begin{figure}[h!]
    \centering
    \includegraphics[width=\linewidth]{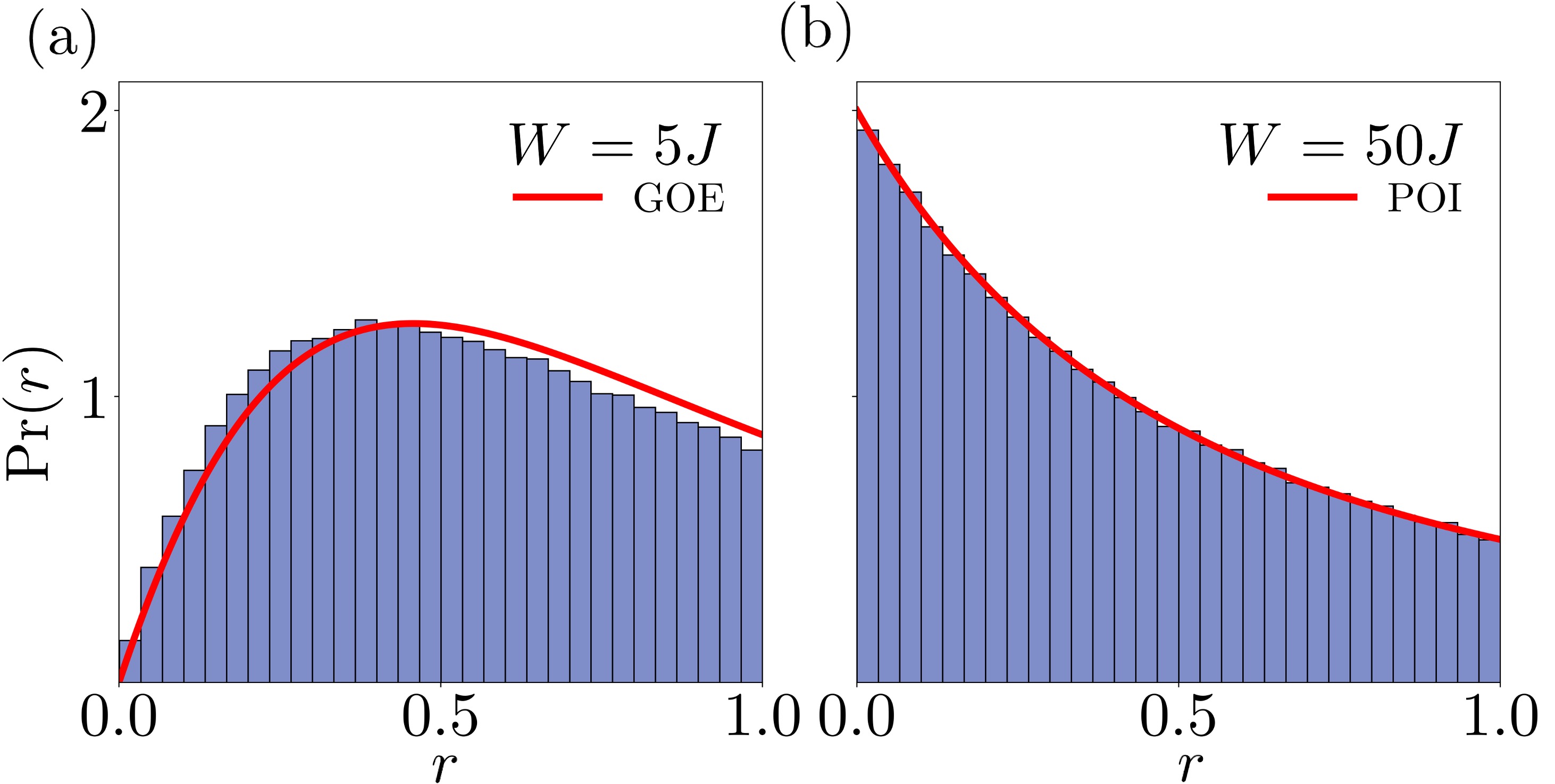}
    \caption{\textbf{Level statistics of the disordered Ising model.} The histograms display the level statistics of 500 instances of 9-qubit systems governed by the Hamiltonian in Eq.~\eqref{eq:IsingModel} for two different disorder strengths. (a) With $W=5J$, the histogram follows the GOE distribution, indicating that the system is in the thermalized phase characterized by level repulsion and correlated eigenvalues. (b) With $W=50J$, the histogram follows the Poisson distribution, indicating that the system is in the MBL phase where eigenvalues are uncorrelated and level crossings are allowed. These different disorder strengths are used to initialise the parameters in our ans\"{a}tze for the thermalized and MBL initialisations.}
    \label{fig:level_stat}
\end{figure}

In Fig.~\ref{fig:level_stat}, we illustrate the two different phases exhibited by the analog model in Eq.~\eqref{eq:IsingModel} using the level statistics (with $W=5J$ for the thermalized phase, and $W=50J$ for the MBL phase; both with $B = -2J$). Note that in the limit where the disorder is absent $W \approx 0$, the model in Eq.~\eqref{eq:IsingModel} becomes equivalent to a system of non-interacting fermions and hence departing from the thermalized phase~\cite{russomanno2016entanglement}. We refer to Appendix~\ref{appx:level_stat} for more details of the level statistics with varying disorder strengths.

\subsection{Initialisation}
To associate the notion of quantum phases into our analog VQA framework, we consider two initialisation strategies corresponding to the thermalized and MBL phases. First, the {\bf thermalized initialisation} involves setting each quench in Eq.~\eqref{eq:quench_dynamics} to be in the thermalized phase. Specifically, for each quench, the variational parameters are independently drawn with a disorder strength $W$ such that $H(\thv_m)$ corresponds to the thermalized phase. Similarly, the {\bf MBL initialisation} is defined by choosing the disorder strength $W$ for each quench such that $H(\thv_m)$ corresponds to the MBL phase at long times. We note that in these strategies, we enforce that all quenches correspond to the same phase (either all thermalized or all MBL), but the parameters  $\thv_{m}$ for different quenches are generally different, i.e., $\thv_{m} \neq \thv_{m'}$ for $m \neq m'$.

\section{Main results}
\label{main-result}

We study two key fundamental aspects of VQAs in the context of the analog simulators, namely expressivity and the emergence of BP phenomenon and connect them to the phases of matter. The summary of our fundamental results is illustrated in Fig.~\ref{fig:summary}. In addition, we propose the practical BP-free initialisation strategy based on MBL phase.

\subsection{Expressivity}
\label{subsec:expressivity}

The expressivity of a model considered in this work refers to its ability to uniformly explore the entire Hilbert space. More precisely, we consider the unitary ensemble of the parametrized ans\"{a}tze corresponding to each phase, defined as 
\begin{align}\label{eq:ansatze_ensemble}
    \mathbb{U}_{\Theta_{D}} = \{ U(\thv) \}_{\thv \in \Theta_{D}} \;,
\end{align}
where $\Theta_D$ is a region of the parameter space such that the Hamiltonian in each quench is in the corresponding phase $D$ (here, $D = {\rm ``Thermalized"}$ or $D = {\rm ``MBL"}$). This expressivity measures how close the ensemble $\mathbb{U}_{\thv_D}$ is to forming a unitary 2-design over the unitary group -- that is, a pseudo-random distribution that matches the Haar random distribution up to the second moment~\cite{sim2019expressibility, holmes2021connecting, thanasilp2022exponential}. 

To formalize this, let us consider the $t$-order frame potential~\cite{sim2019expressibility} 
\begin{align}\label{eq:frame_poten}
\FC^{(t)}_{\mathbb{U}_{\Theta_{D}}} = \int\int_{U,V\in \mathbb{U}_{\Theta_{D}}} d\mu(U)d\mu(V) \left|\langle 0|V^\dagger U|0\rangle \right|^{2t} \;,
\end{align}
where $d\mu(U)$ (and $d\mu(V)$) is a probability measure of $U$ (and of $V$) over the ensemble $\mathbb{U}_{\Theta_{D}}$. The expressivity can be quantified by the second-order frame potential $\FC^{(2)}_{\mathbb{U}_{\Theta_{D}}}$, which reaches its minimum when the ensemble forms a unitary $2$-design, in which case $\FC_{\rm Haar}^{(2)} = 2(2^n-1)!/(2^n+1)!$. Because a 2-design is automatically a 1-design, the same ensemble also saturates the Haar value
$\mathcal{F}^{(1)}_{\rm Haar}=1/2^{\,n}$. Note that there also are alternatives for defining expressivity in VQAs, such as through output function space~\cite{perez2020data, schuld2021effect,gil2024expressivity,gan2023unified}, or geometric structure of the parametrized states~\cite{haug2021capacity, abbas2020power, larocca2021theory}.

\begin{figure}[!h]
    \centering
    \includegraphics[width=\linewidth]{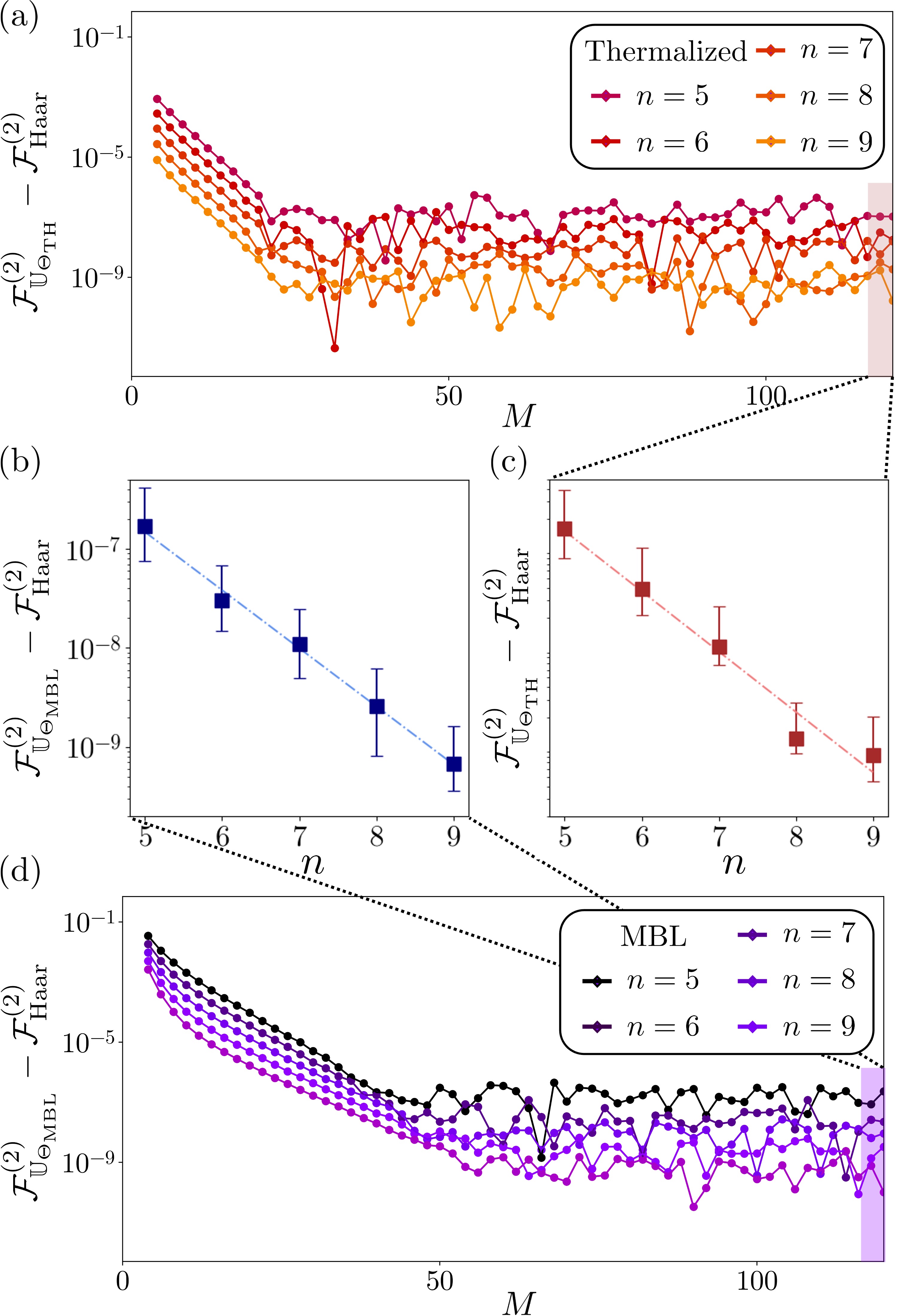}
    \caption{\textbf{Second-order frame potentials} (a) The difference between the second-order frame potential of the ans\"{a}tze ensemble and of the Haar distribution is plotted against the number of quenches $M$ for the systems size ranging from $n=5$ to $n=9$, under the thermalized initialisation. The second-order frame potential is estimated by averaging the square of fidelity over $N(N-1)/2$ independent pairs with $N=30,000$. The saturated values of the second-order frame potential difference are plotted against the system size $n$ for (b) the thermalized and (c) MBL initialisations. (d) Same as panel (a), but under the MBL initialisation.} 
    \label{fig:2FP}
\end{figure}

Quench dynamics drive the ensemble toward a 2‑design as the number of quenches $M$ grows, and we therefore expect both phases to attain maximal expressivity at large $M$.  Figure~\ref{fig:2FP} confirms this expectation numerically, with the thermalized phase converging markedly faster than the MBL phase, consistent with its more chaotic dynamics and the larger
portion of Hilbert space explored by each thermalized quench.

\subsection{Barren plateaus}
\label{sec:BP}
\begin{figure*}[htbp]
    \centering
    \includegraphics[width=\textwidth]{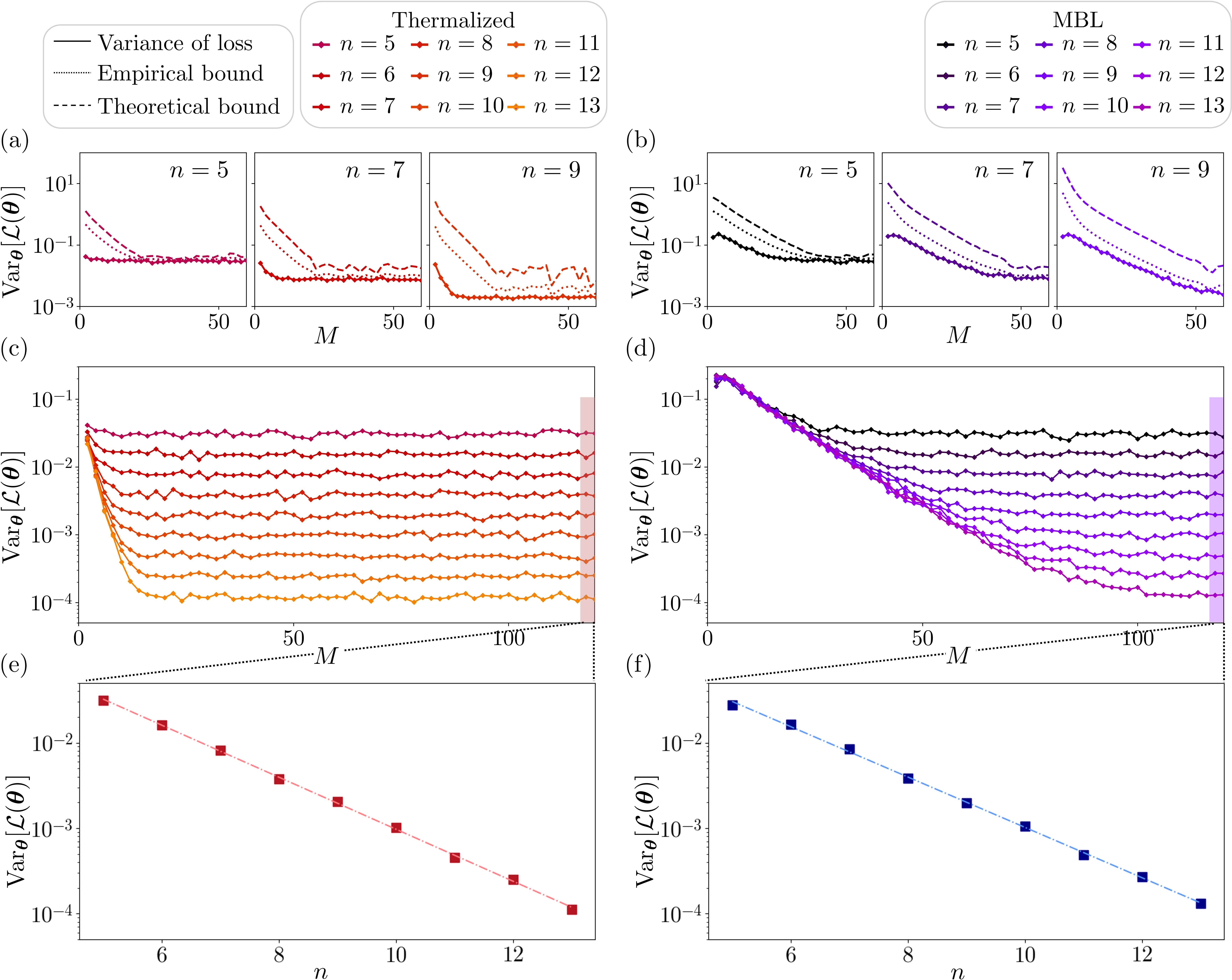}
     \caption{\textbf{Variance of the loss function in the thermalized and MBL initialisations.} The top panels illustrate the comparison between the variance of the loss function for (a) thermalized and (b) MBL initialisations and its bounds against the number of quenches $M$. Specifically, the bounds are functions of the frame potential difference between the ans\"{a}tze ensemble and the Haar distribution. The solid line represents the variance of the loss $\langle Z_1Z_2\rangle_{\thv}$, the dotted line shows the empirical bound for the variance, and the dashed line indicates the theoretical bound for the variance as presented in Eq.~\eqref{eq:variance-expressivity-bound}. These panels display data for 5, 7, and 9 qubits from left to right. The middle panels show the variance of the loss $\langle Z_1Z_2\rangle_{\thv}$ for (c) thermalized and (d) MBL initialisation as the number of quenches increases for systems with 5 to 13 qubits, averaged over 400 realizations. The bottom panels present the saturated variance of $\langle Z_1Z_2\rangle_{\thv}$ plotted on a logarithmic scale against the number of qubits for (e) thermalized and (f) MBL phase. This provides evidence for the possible emergence of barren plateaus when the ans\"{a}tze is initialised in both phases.}
    \label{fig:BP}
\end{figure*}

One challenge that prevents the scalability of VQAs is the BP phenomenon, where the loss landscape on average becomes exponentially flat with increasing system size~\cite{mcclean2018barren, larocca2024review,cerezo2025does,sharma2020trainability,arrasmith2021equivalence,larocca2021diagnosing,holmes2021connecting,khatri2019quantum,rudolph2023trainability,kieferova2021quantum,thanaslip2021subtleties,tangpanitanon2020expressibility,holmes2021barren,martin2022barren,fontana2023theadjoint,ragone2023unified, thanasilp2022exponential, letcher2023tight,anschuetz2024unified, chang2024latent, deshpande2024dynamic,crognaletti2024estimates,  mao2023barren,mhiri2025unifying,cerezo2021cost,uvarov2020barren,marrero2020entanglement,patti2021entanglement,wang2020noise,mele2024noise,xiong2023fundamental,xiong2025role}. Upon randomly initialising the training over the BP region, the number of measurement shots required from quantum hardware to reliably navigate through the landscape also scales exponentially with the system size. This renders training VQAs impractical for large numbers of qubits. 

More formally, we say the loss landscape is exponentially flat over the region of the loss landscape with parameters sampled according to probability i.e., $\thv \sim \mathcal{P}$ 
if
\begin{align}
    {\rm Var}_{\thv \sim \mathcal{P}} [\LC(\thv)] \in \OC\left( \frac{1}{b^n} \right) \;,
\end{align}
for some $b>1$, where $n$ is the number of qubits. 

Intuitively, the root of BPs can be understood from \textit{the curse of dimensionality} perspective, where parametrized quantum states in an exponentially large Hilbert space are poorly handled~\cite{cerezo2025does}. Various sources that lead to the occurrence of BPs due to these inappropriately designed parametrized states have been identified, including excessive expressivity in the ans\"{a}tze~\cite{mcclean2018barren,holmes2021connecting,larocca2021diagnosing}, global measurements~\cite{uvarov2020barren,cerezo2021cost}, high levels of entanglement~\cite{marrero2020entanglement,patti2021entanglement}, and the presence of noise~\cite{wang2020noise,mele2024noise}. In addition, while initially discussed in the context of VQA untrainability, the curse of dimensionality is later found to also plague the scalability of non-variational quantum models such as quantum kernel methods~\cite{thanasilp2022exponential, kubler2021inductive, shaydulin2021importance, huang2021power,suzuki2023effect, suzuki2024quantum} and quantum reservoir processing~\cite{xiong2023fundamental,xiong2025role}. Furthermore, an increasing number of approaches have been proposed to tackle BPs~\cite{gelman2024survey,larocca2024review} including expressivity-limited architectures~\cite{volkoff2021large,skolik2021layerwise}, imposing symmetries on circuits~\cite{gard2020efficient,zhang2021shallow,lyu2023symmetry,meyer2023exploiting}, as well as employing alternative initialisation strategies~\cite{verdon2019learning,grant2019initialization,lyu2020accelerated,patti2021entanglement,zhang2022escaping,liu2023mitigating,park2024hardware,cao2024exploiting,xin2024improve}. Lastly, caution must be exercised when claiming that an approach can circumvent BPs, particularly due to the complex interplay with shot noise~\cite{aghaei2025pitfalls}.

In this work, we mainly focus on studying BPs arising from the expressivity and entanglement of the analog quench dynamics and their connection to the quantum phases in which the ans\"{a}tze operates. Following similar treatments as in Ref.~\cite{holmes2021connecting} which studies the relationship between expressivity and BPs, we have the loss variance bound at the level of loss values expressed as
\begin{align}\label{eq:variance-expressivity-bound}
    \Var_{\thv\sim \PC}[\LC(\thv)] \leq \Var_{\rm Haar}[\LC(\thv)]  +\BC_n\big(\mathbb{U}_{\Theta_{D}},O\big) \;\;,
\end{align}
with the expressivity-dependent correction term
\begin{align}
\BC_n\big(\mathbb{U}_{\Theta_{D}},O\big) = &\left( \sqrt{\FC^{(2)}_{\mathbb{U}_{\Theta_{D}}} - \FC^{(2)}_{\rm Haar}}-\FC^{(1)}_{\mathbb{U}_{\Theta_{D}}} + \FC^{(1)}_{\rm Haar}\right) \| O\|^2_2 \nonumber\\
&+ \frac{\Tr[O]}{2^{n-1}} \sqrt{\FC^{(1)}_{\mathbb{U}_{\Theta_{D}}} - \FC^{(1)}_{\rm Haar}} \| O\|_2.  
\end{align}
Remark that the bound in Eq.~\eqref{eq:variance-expressivity-bound} is saturated as the ensemble approaches a unitary 2-design
$(\mathcal{F}^{(2)}_{\mathbb{U}_{\Theta_D}}\to\mathcal{F}^{(2)}_{\rm Haar}$). Namely,
the correction term $\mathcal{B}_{n}$ vanishes and the variance
saturates at its Haar value. The proof of Eq.~\eqref{eq:variance-expressivity-bound} is presented in Appendix~\ref{appx:proof_bound}.

\begin{figure}[!ht]
    \centering
    \includegraphics[width=\linewidth]{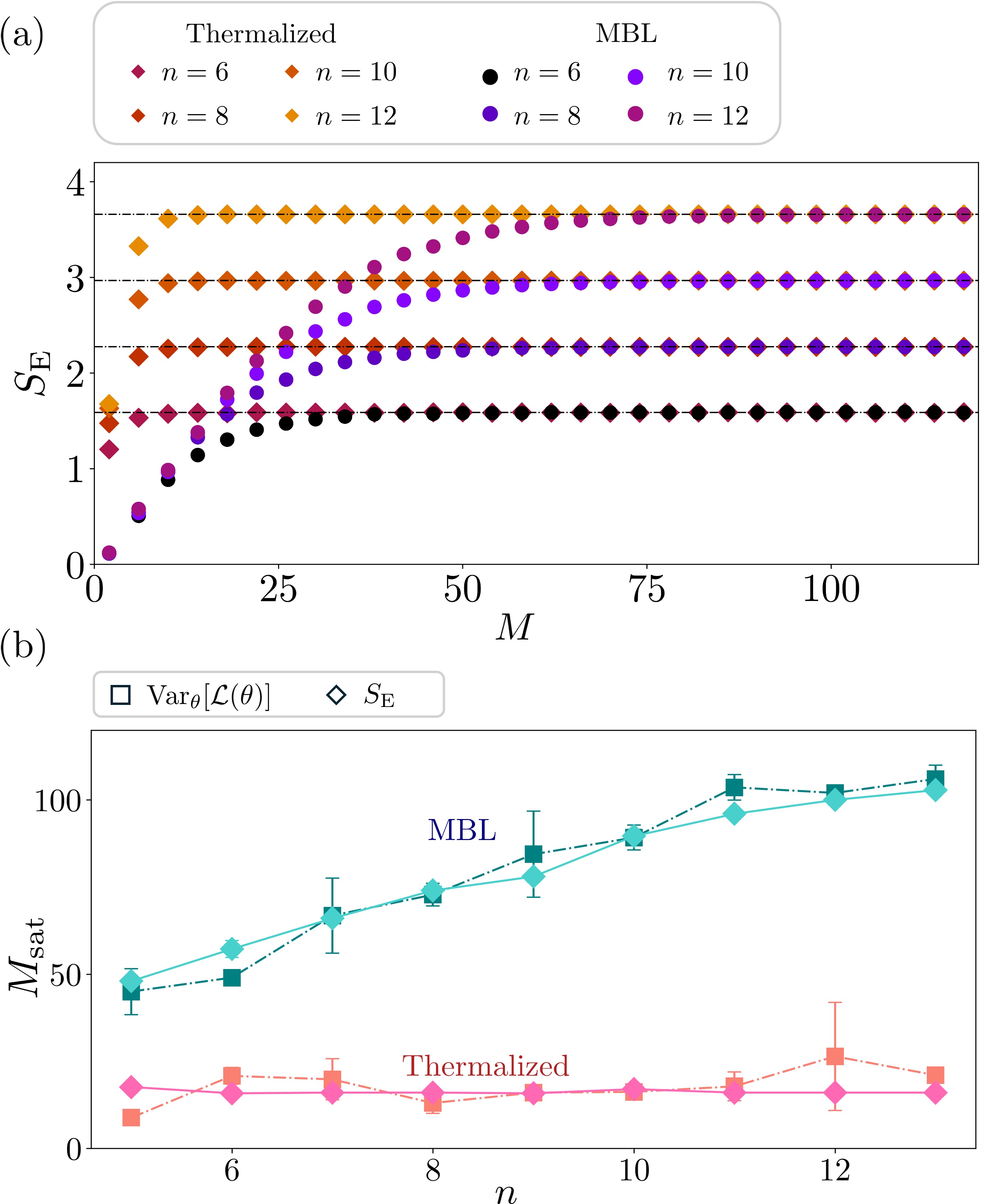}
    \caption {\textbf{Entanglement–trainability correspondence for thermalized and MBL initialisation} (a) Bipartite  von Neumann entanglement entropy $S_{\rm E}$, averaged over 400 realizations, is plotted against the number of quenches $M$ for a bipartition made at the middle of the chain. Results are shown for the thermalized phase (Diamond) and MBL phase (Circle). The horizontal dotted lines indicate the average bipartite entanglement entropy for Haar-random states. The faster convergence of the thermalized initialisation compared to the MBL initialisation is attributed to the differences in entanglement growth characteristics of each phase. (b) The numbers of quenches required for the convergence, $M_{\rm sat}$, of the variance of $\expval{Z_1Z_2}_{\thv}$ (Square) and the entanglement entropy (Diamond) are plotted against the number of qubits for both the thermalized and the MBL initialisations. The consistent trend between entanglement and the onset of barren plateaus in both phases highlights the role of quantum phases in shaping the untrainability of the analog VQA ans\"{a}tze.}
    \label{fig:EE}
\end{figure}

In Fig.~\ref{fig:BP}, we study the variance scaling of the local loss in different phases with respect to the number of qubits and the number of quenches. The observable is chosen to be a product of Pauli-Z operators on the first and second qubits i.e., $\LC(\thv) = \langle Z_1 Z_2 \rangle_{\thv}$, to ensure that the observed BPs do not originate from global measurements~\cite{cerezo2021cost}. 

Crucially, the overall trend in both phases aligns with what we observe from the expressivity scaling in Fig.~\ref{fig:BP}. Firstly, in panels (a) and (b), we verify Eq.~\eqref{eq:variance-expressivity-bound} by explicitly plotting the upper bounds for different quenches and for some fixed qubits. In addition, we also found an empirical bound of the form \begin{align}
\widetilde{\BC}_n\big(\mathbb{U}_{\Theta_{D}},O\big) = \,&\left( \sqrt{\FC^{(2)}_{\mathbb{U}_{\Theta_{D}}} - \FC^{(2)}_{\rm Haar}}-\FC^{(1)}_{\mathbb{U}_{\Theta_{D}}} + \FC^{(1)}_{\rm Haar}\right) \| O\|^{2k}_2 \nonumber \\
&+ \frac{\Tr[O]}{2^{n-1}} \sqrt{\FC^{(1)}_{\mathbb{U}_{\Theta_{D}}} - \FC^{(1)}_{\rm Haar}} \| O\|_2 
\end{align}
with $k=0.7$, which is tighter than the theoretical one with $\BC_n\big(\mathbb{U}_{\Theta_{D}},O\big)$, suggesting that there is room for  improvement on the tightness of the expressivity bound. Next, as presented in panels (c) and (d), for a large number of quenches, the loss variances in both phases saturate at the same value for a fixed number of qubits. When plotting the saturated values against the number of qubits (as shown in the panels (e) and (f)), we observe the exponential vanishing of the variance, demonstrating the presence of BPs around the initialisation region. 
Furthermore, the decay rate of the variance in the thermalized phase is observed to be much faster than that in the MBL phase.

The flatness of the landscape can also be physically understood from the perspective of entanglement growth in each phase. In particular, for each thermalized quench, the amount of entanglement generated is much larger than that in the MBL case. In panel (a) of Fig.~\ref{fig:EE}, we plot the growth of entanglement entropy as a function of the number of quenches for different phases. 
One can see that the convergence behavior of the entanglement entropy, shown in Fig.~\ref{fig:EE}, mirrors that of the local observable variance in  Fig~\ref{fig:BP}. 

Lastly, as shown in panel (b) of Fig.~\ref{fig:EE}, the number of quenches $M_{\rm sat}$ corresponding to the saturation values of the loss variance, and entanglement entropy share similar scaling behavior with respect to the system size. In particular, the gap between the number of quenches required for saturation in the thermalized and MBL phases is empirically observed to increase linearly with the number of qubits.

\subsection{Initialisation strategy}
\label{sec:strategy}

\begin{figure}[!ht]
    \centering
    \includegraphics[width=\linewidth]{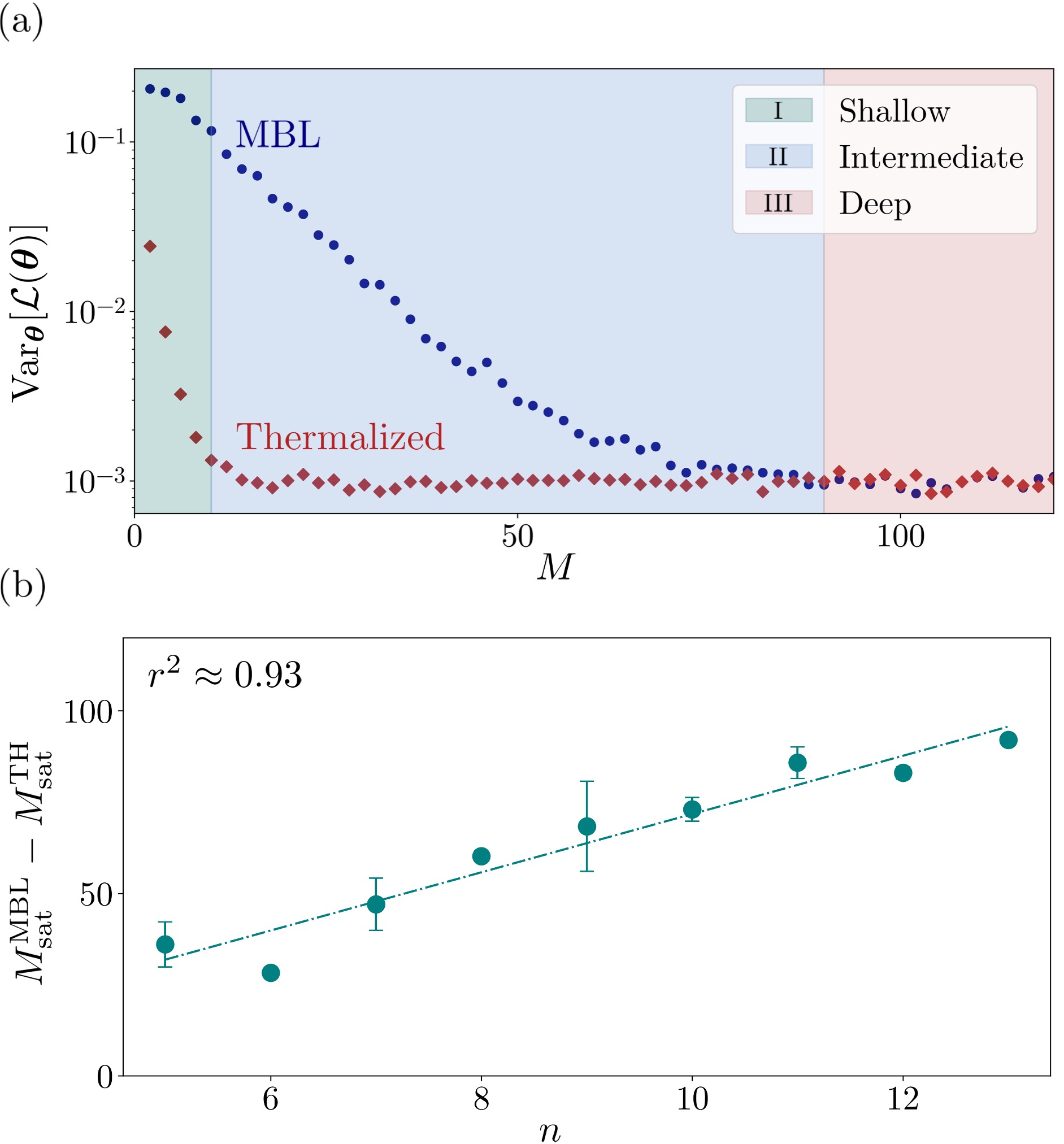}
    \caption {\textbf{Three initialisation regimes.} (a) The variances of the local loss function  for the thermalized (Diamond) and MBL (Circle) initialisations are plotted against the number of quenches for a 10-qubit system. To categorize the initialisation regimes, we divide them into three based on the differences in variance between the thermalized and MBL initialisations: {\bf Regime I (Shallow)} (Green) captures the range of quenches where both initialisations are trainable and do not exhibit barren plateaus, {\bf Regime II (Intermediate)} (Blue) defines  the range of quenches where the MBL initialisation remains trainable, while the thermalized counterpart already has suffered from the barren plateau, and {\bf Regime III (Deep)} (Red) encapsulates the range of quenches where both phases exhibit barren plateaus, making the VQA untrainable. (b) The difference in the number of quences between the MBL and the thermalized initialisation when BP onset is plotted against the number of qubits. This difference is the width of Regime II. As the number of qubits increases, the difference grows linearly with the number of qubits. This suggests the advantage of the MBL initialisation over the thermalized initialisation.
    }
    \label{fig:initializing_regime}
\end{figure}

From the fundamental properties of the analog quench ans\"{a}tze empirically studied in the previous section, we propose an initialisation strategy that provides polynomially large gradients at the initial stage of training while leaving room for large expressivity at a later stage of training. To see how this is possible, we note that one can categorize the initialisation regimes based on the number of quenches at which the saturation onset occurs in each phase as follows:
\begin{itemize}
    \item Regime I (\textit{small} number of quenches):  Before the saturation in thermalized initialisation, both thermalized and MBL phases are free from barren plateaus but lack sufficient parameters to reach maximum expressivity.
    \item Regime II (\textit{intermediate} number of quenches): After regime I, the thermalized phase becomes maximally expressive but suffer from barren plateaus. In contrast, the MBL phase exhibits a large initial loss difference and becomes increasingly expressive compared to the Regime I.
    \item Regime III (\textit{large} number of quenches): Once the MBL initialisation loss value saturates, both phases are maximally expressive but enter barren plateau regime.
\end{itemize}
This categorization is summarized in Fig.~\ref{fig:summary} and numerically illustrated in the top panel of Fig. \ref{fig:initializing_regime}.

In practice, employing an ans\"{a}tze for a standard VQAs typically requires predetermining the number of quenches (i.e., treating $M$ as a hyperparameter). 
Hence, our findings suggest a novel initialisation strategy where we initialise the ans\"{a}tze in the MBL phase when the number of quenches falls within Regime~II. With this strategy, the benefits are twofold. First, the initial training is guaranteed not to suffer from BPs. Second, during later training iterations, the parameters can be tuned such that the ans\"{a}tze no longer corresponds to the MBL phase. In this scenario, the number of quenches in Regime~II is sufficient in the sense that  maximal expressivity can, in principle, be achieved by the ans\"{a}tze as indicated by the saturation in the thermalized phase. Additionally, we note that by parameterizing dynamic times and interaction strengths for individual quenches, the ans\"{a}tze becomes universal, being able to express any unitary~\cite{parra2020digital,wiersema2023classification,hu2025universal}. In the next Section~\ref{sec:demonstration}, as well as Appendix~\ref{appx:extended_numerical_result}, we benchmark the proposed MBL initialisation scheme on small toy model examples (up to $10$ qubits), including a ground state finding task and a Max-Cut problem.

One could then further ask whether this initialisation strategy is scalable with the system size. As illustrated in Fig.~\ref{fig:initializing_regime}, the difference between the least number of quenches required for the MBL and thermalized phases to achieve  saturated variance is empirically found to scale linearly with the number of qubits. Hence, this suggests that Regime~II should persist for large system sizes.   

While our strategy offers a promising approach to mitigate barren plateaus, it is essential to stress that it does not entirely avoid them. More precisely, the ans\"{a}tze with the number of quenches in Regime~II is sufficiently expressive such that there are BP regions in the loss landscape.  While the training is initialised in a region with substantial gradients, there is no guarantee that the training trajectory will not later wander into flat regions. Additionally, this initialisation strategy can be seen as an analog version of the close-to-identity initialisation \cite{grant2019initialization, zhang2022escaping}, which is generally regarded as problem-agnostic. Hence, there is no guarantee that the solution obtained  using this strategy will correspond to a good local minimum.

\subsection{Numerical results}
\label{sec:demonstration}
To demonstrate the capability of the proposed MBL initialisation scheme, we apply it to solve two standard optimization problems: (i)~Finding the ground state energy of some target Hamiltonians, and (ii)~Solving a combinatorial optimization. We compare the performance of the MBL initialisation strategy against the thermalization-based strategy. The depths for each are chosen such that both ansätze exhibit comparable variance in the loss landscape: a shallow depth for the thermalized phase and an intermediate depth for the MBL phase.

Specifically, we employ a shallow depth of $M=2$ quench for the thermalized initialisation, and an intermediate depth of $M=22$ (for $n=6$) or $M=24$ (for $n=8$) for the MBL initialisation. Although the initialisation phases are fixed, during the optimization process we train all relevant parameters of the ansätze, including the longitudinal field, the transverse field, and the evolution time for each quench.

\subsubsection{Finding the ground state of some Hamiltonian}
We test our approach by considering two target Hamiltonians in an $8-$qubit system:
\begin{itemize}
\item \textbf{Long-range disordered Ising model:} A disordered Ising model with power-law decaying long-range interactions is of the form
\begin{equation}
    \label{eq:ATA_ising}
       H_{\rm Ising} = J\sum_{i<j}\frac{Z_iZ_j}{|i-j|^\alpha}+ \sum_{i=1}^n h_iX_i\;\;,
\end{equation}
where we choose $\alpha=1$ and the local fields $h_i$ are sampled independently from a uniform distribution over $[-0.3J,0.3J]$. 
\item \textbf{Heisenberg model:} The Heisenberg model with periodic boundary condition is defined as
\begin{equation}
\label{eq:Heisenberg_XXX}
H_{\rm XYZ}=J\sum_{i=1}^n(X_iX_{i+1}+Y_iY_{i+1}+Z_iZ_{i+1})\;\;.
\end{equation}
\end{itemize}

The performance is evaluated using the relative error, defined as the normalized difference between the energy obtained at a particular iteration $E$
and the true ground state energy $E_\text{min}$:
\begin{equation}
    \Delta E = \frac{|E -E_\text{min}|}{E_\text{min}}\times100\%\;\;.
\end{equation}

\begin{figure}[!ht]
    \centering
    \includegraphics[width=\linewidth]{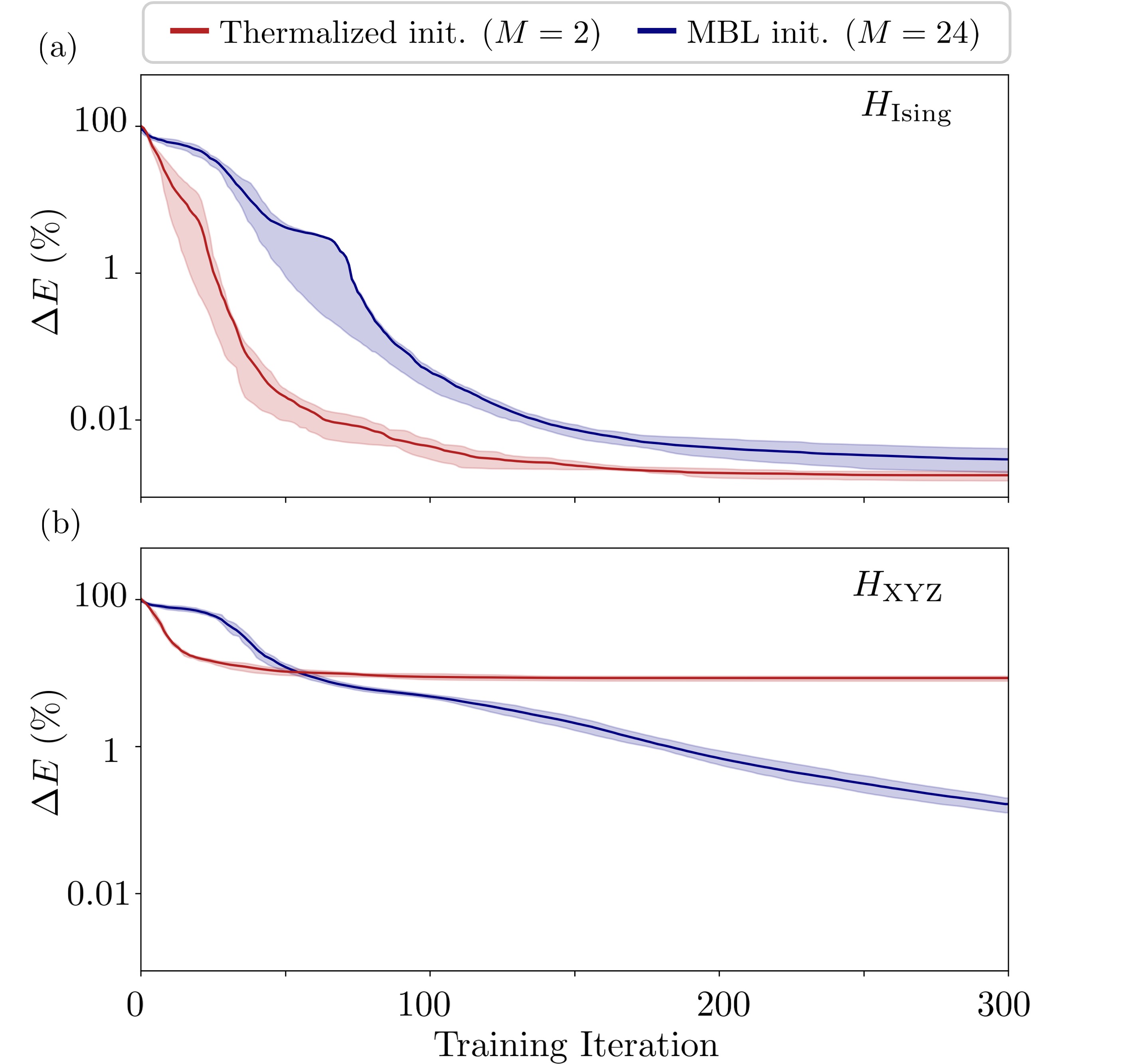}
    \caption {\textbf{Ground state estimation results.} The average relative error over $5$ initialisations is plotted against training iterations on a logarithmic scale for estimating two target Hamiltonians in the $8$-qubit systems: (a) the long-range disordered Ising model (Eq.~\ref{eq:ATA_ising}) and (b) the Heisenberg model with periodic boundary conditions (Eq.~\ref{eq:Heisenberg_XXX}). The red line represents the optimization with thermalized initialisation for $M=2$, while the blue line shows the optimization with the MBL initialisation for $M=24$. The shaded region around each line indicates the $20^\text{th}$ to $80^\text{th}$ percentile range of the relative error. } 
    \label{fig:VQE_result}
\end{figure}

Figure~\ref{fig:VQE_result} shows the results, averaged over 5 initialisations after 300 training iterations. For the case of $H_{\rm Ising}$, 
both initialisation strategies achieve comparable performance with 
an averaged relative error of $1.44\times10^{-3}\%$ for the thermalized initialisation and $1.82\times10^{-3}\%$ for the MBL initialisation. 
We argue that since the target Hamiltonian $H_{\rm Ising}$ structurally resembles the ansätze Hamiltonian in Eq.~\eqref{eq:IsingModel}, the ansätze possesses a built-in \emph{inductive bias} towards the solution.  This facilitates the faster convergence of the thermalized initialisation strategy despite its limited expressivity at shallow quenches.

Next, we discuss the results for the Heisenberg model $H_{\rm XYZ}$. Here, the ansätze is less structurally aligned with the solution of the form $H_{\rm XYZ}$.Empirical observations show that the MBL initialisation achieves an averaged relative error of $0.11\%$, significantly outperforming the thermalized initialisation, which yields a $7.7\%$ error.
We attribute this performance discrepancy to the enhanced expressivity of the MBL initialisation. Consequently, for target Hamiltonians that are less structurally aligned with the ansätze, the MBL initialisation in the intermediate quench regime is a more suitable candidate.

\subsubsection{Combinatorial optimization}
Here we consider the Max-Cut problem, a canonical combinatorial optimization task~\cite{Haribara2016coherentising}. The problem asks for a partition of a graph's vertices into two disjoint sets such that the sum of weights of edges connecting the two sets is maximized. Despite its simple definition, Max-Cut is NP-hard on general graphs~\cite{Haribara2016coherentising,lucas2014ising}.

To solve this using a quantum framework, we map the problem to finding the ground state of an Ising spin glass. By assigning a spin $Z_i = \pm 1$ to each vertex, the cut size is maximized when connected spins are anti-aligned (i.e., $Z_i Z_j = -1$). This is equivalent to minimizing the Ising Hamiltonian~\cite{lucas2014ising}:
\begin{equation}
    \label{eq:H_MC}
    H_{\rm MC} = \sum_{i <j} w_{ij} Z_i Z_j \;\;,
\end{equation}
where $w_{ij}$ are the edge weights defined by the problem instance. We focus on a family of random complete graphs where the connectivity is all-to-all and the weights $w_{ij}$ are sampled uniformly from $[-1,1]$. This fully connected model with random weights corresponds to the Sherrington-Kirkpatrick spin glass model, which is known to possess a complex energy landscape with many local minima.

The optimization objective is to maximize the cut value, which relates to the Hamiltonian via the linear transformation $C(\vec{\theta}) \propto \text{const} - \langle H_{\rm MC} \rangle$. Specifically, the cost function is defined as~\cite{Haribara2016coherentising, lucas2014ising}:
\begin{equation}
\label{eq:max-cut_cost}
    C(\vec{\theta}) = \frac{1}{2}\sum_{i<j} w_{ij} (1 - \langle Z_i Z_j \rangle_{\vec{\theta}}) = \frac{1}{2}\left(\sum_{i<j}w_{ij}-\expval{H_{\rm MC}}_{\vec{\theta}}\right) \;\;.
\end{equation}
To evaluate the solution quality, we use the approximation ratio
\begin{equation}
    \text{Approx. Ratio} = \frac{C(\vec{\theta})}{C_{\text{max}}}\times100 \% \;\;,
\end{equation}
which measures how close the obtained cost value (ansätze-dependent) is to the global maximum $C_{\text{max}}$, the ground state of the target problem (ansätze-independent)~\cite{abbas2024challenges}.

\begin{figure}[!ht]
    \centering
    \includegraphics[width=\linewidth]{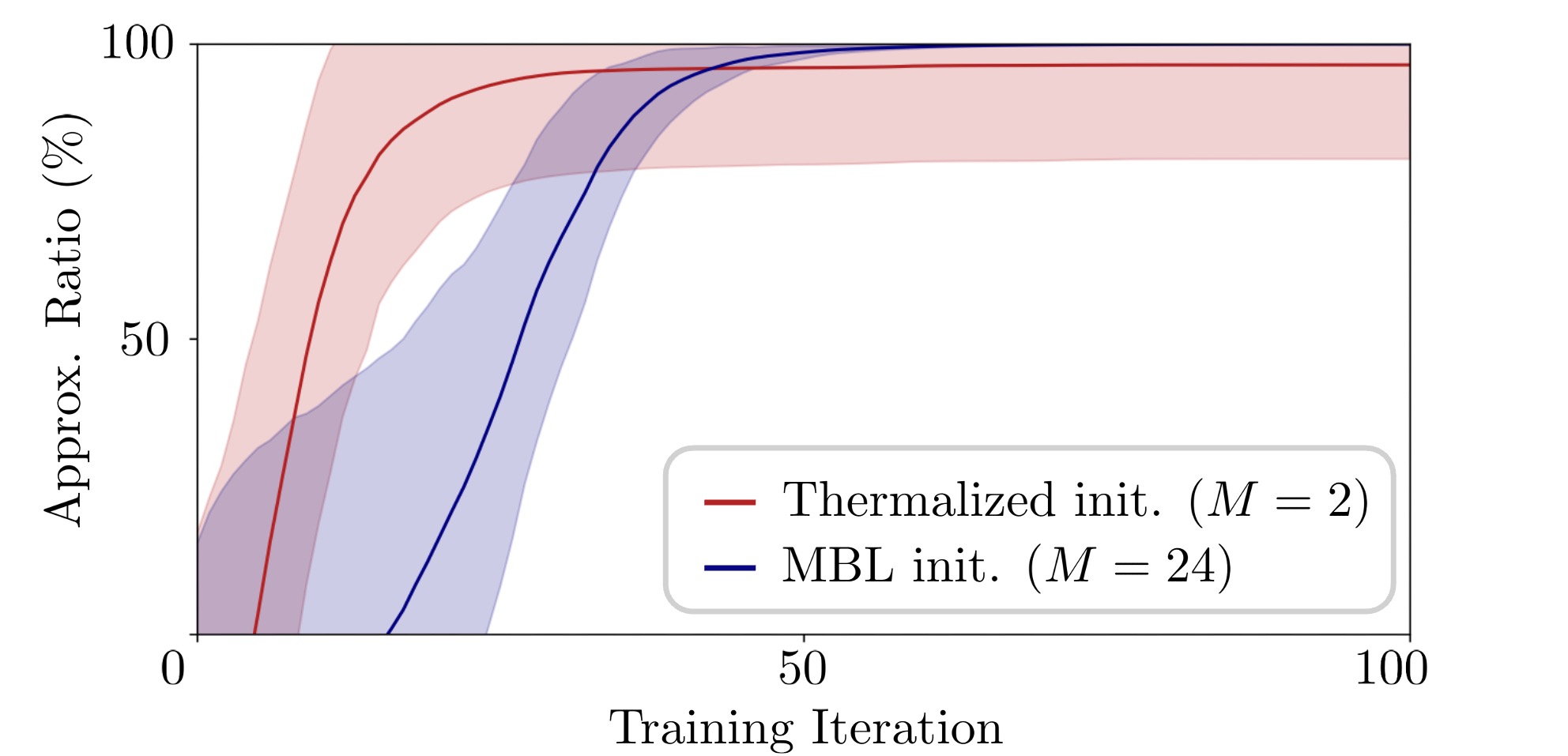}
    \caption {\textbf{Max-Cut optimization results.}  The average approximation ratio of the Max-Cut problem on $10$ random weighted complete graphs with $n=6$ vertices is plotted against training iterations. Results are averaged over $5$ initialisation per graph instance. The red line represents the initialisation in the thermalized phase with $M=2$, and the blue line represents the MBL phase with $M=22$. The shaded area around each line indicates the standard deviation of the approximation ratio.}
    \label{fig:QUBO_result}
\end{figure}    

Figure~\ref{fig:QUBO_result} shows the optimization result after 100 training iterations for graphs with $6$ vertices, averaged over 10 random graph instances and 5 initialisations (for each strategy) per instance. The thermalized initialisation strategy achieves an average approximation ratio of $96.5\%$ 
, which is slightly lower than the MBL initialisation's ratio of $99.9\%$. 

Furthermore, we empirically observe that while the shallow  quenches initialised in the thermalized phase has sufficient expressivity to represent the ground state, the training occasionally becomes trapped in local minima, as indicated by the relatively large shaded variance region. On the other hand, the MBL initialisation strategy shows a reliable convergence to the ground state. This observation can be explained by the overparameterization phenomenon~\cite{larocca2021theory}. In particular, the number of parameters used for $M=22$ in the $n=6$ case significantly exceeds the dimension of the Dynamical Lie Algebra of our ansätze~\cite{wiersema2023classification}, which satisfies the lower bound required for the overparametrized regime~\cite{larocca2021theory}. We note, however, that for larger system sizes,  exponentially many parameters cannot be practically realized; hence, the interplay between the MBL initialisation strategy in the intermediate regime and the local minima traps  remains to be explored. We refer readers to Appendix~\ref{appx:extended_numerical_result} for extended results.

\section{Discussion and Outlook}
\label{sec:Discussion}
This work proposes a series of analog quench dynamics as an ans\"{a}tze for variational quantum algorithms (VQAs) and investigates the interplay between quantum phases in which the ans\"{a}tze operates and fundamental aspects of VQAs, including expressivity and the occurrence of barren plateaus (BPs). Our findings indicate that the dynamical properties of each quantum phase influences how fast expressivity increases and how quickly the loss landscape becomes flat. Specifically, the chaotic dynamics induced by the thermalized phase ans\"{a}tze lead to maximal expressivity and the onset of BPs at intermediate numbers of quenches, while the many-body localized (MBL) phase ans\"{a}tze reaches this regime much later due to the slower growth of entanglement. Overall, the trends of expressivity increases and variance scaling of the loss landscape closely follow each other, in agreement with the results found in Ref.~\cite{holmes2021connecting}. 

Based on these observations, we propose an initialisation strategy where the ans\"{a}tze operates in the MBL phase with the number of quenches chosen such that the MBL phase does not yet suffer from BPs and the thermalized phase has already achieved the maximal expressivity. 
This approach aims to maintain significant geometric features in the loss landscape around the initial training point, thereby preserving initial trainability, while ensuring sufficient expressivity for later training iterations. We note however that although this initialisation strategy can help to circumvent BPs, it is generally problem-agnostic. Consequently, the quality of the solution obtained using this strategy cannot be universally guaranteed, and its effectiveness may vary depending on the specific problem or objective function, as shown in the optimization results for VQE problems and a Max-Cut problem in Section~\ref{sec:demonstration} and Appendix~\ref{appx:extended_numerical_result}.

On a related note, recent studies suggest a link between the absence of BPs and the classical surrogatability of the loss landscape~\cite{cerezo2025does}. By allowing an initial data acquisition phase on quantum hardware with non-classical initial states, it may be possible to efficiently reconstruct a BP-free loss landscape for classical optimization~\cite{mhiri2025unifying, puig2024variational,lerch2024efficient,pesah2020absence, bermejo2024quantum, goh2023lie, schatzki2022theoretical}. In this aspect, the MBL initialisation in our analog ans\"{a}tze is expected to be consistent with these claims.
Investigating this relationship in details requires both theoretically proving polynomial scaling of the loss variance which can perhaps be achieved by using the framework for generic alternative initialisations developed in Ref.~\cite{mhiri2025unifying}, as well as carefully analysing resource scaling for classical simulation with approaches such as tensor-network~\cite{pollmann2016efficient, wahl2017efficient} or Pauli-back propagation methods~\cite{rudolph2023classical,miller2025simulation,rudolph2025pauli}.
 
Moreover, while previous works have utilized MBL concepts to facilitate initialisation strategies that evade BPs~\cite{park2024hardware,cao2024exploiting,xin2024improve}, these efforts have primarily focused on digital gate-based quantum circuits, where the entire circuit initialisation can often be viewed as a single quench of analog MBL dynamics. Here we distinguish our work by focusing on a series of analog quantum dynamics consisting of multiple quenches, thereby providing a different perspective on leveraging MBL phases in VQAs. Our findings are aligned with, and can be seen as an extension of, other MBL initialisation proposals. 

It is crucial to emphasize that BP is not the only challenge facing VQAs. For example, an equally important yet understudied problem is the prevalence of poor local minima in the quantum loss landscape~\cite{kiani2020learning,wiersema2020exploring,anschuetz2021critical,anschuetz2022quantum}. While we observe empirical evidence of this on the small-scale systems, this is far from a comprehensive study. Previous studies of poor local minima~\cite{anschuetz2021critical,anschuetz2022quantum} typically rely on the assumption that the ansätze is constructed from local quantum gates and these results may not directly apply to our analog setting. Further investigation is required to shed the light on  how different quantum phases influence the local minima landscape.

Moving forward, we anticipate that analytical tools such as local integrals of motion (LIOMs)~\cite{serbyn2013local,huse2014phenomenology} could be useful for analytically substantiating the empirical trends observed in our study. Finally, conducting similar investigation on analog models with other different quantum phases such as time-crystals~\cite{wilczek2012quantum,kongkhambut2022observation,zaletel2023colloquium} or topological phases~\cite{wen2017colloquium} is an interesting extension which would deepen our understanding of the role of phases on VQAs. As analog quantum devices continue to advance, implementing and testing experimentally our proposed ans\"{a}tze and initialisation strategy would be a significant step toward practical applications of analog VQAs.

\section{Acknowledgement}
The authors gratefully thank Zo\"{e} Holmes and Marco Cerezo for valuable discussions. 
T.C. and S.T. acknowledge the funding support from the NSRF via the Program Management Unit for Human Resources \& Institutional Development, Research and Innovation [grant number B39G680007].  The authors acknowledge high performance
computing resources including NVIDIA A100
GPUs from Chula Intelligent and Complex Systems
Lab, Faculty of Science, Chulalongkorn
University, Thailand. 
S.T. is partially supported from the Sandoz Family Foundation-Monique de Meuron program for Academic Promotion. 
S.T. further acknowledges the grants for development of new faculty staff, Ratchadaphiseksomphot Fund, Chulalongkorn University [grant number 3230120336 DNS\_68\_052\_2300\_012], as well as funding from National Research Council of Thailand
(NRCT) [grant number N42A680126]. This Research is funded by Thailand Science research and Innovation Fund Chulalongkorn University (IND\_FF\_69\_258\_2300\_062).

\bibliography{quantum.bib, quantum2.bib}

\clearpage
\newpage
\onecolumngrid
\setcounter{theorem}{0}
\setcounter{proposition}{0}
\setcounter{corollary}{0}

\appendix
\vspace{0.5in}
\begin{center}
	{\Large \bf Appendix} 
\end{center}

\section{Characterization of phases via level statistics}
\label{appx:level_stat}

\begin{figure}[h!]
    \centering
    \includegraphics[width=0.9\linewidth]{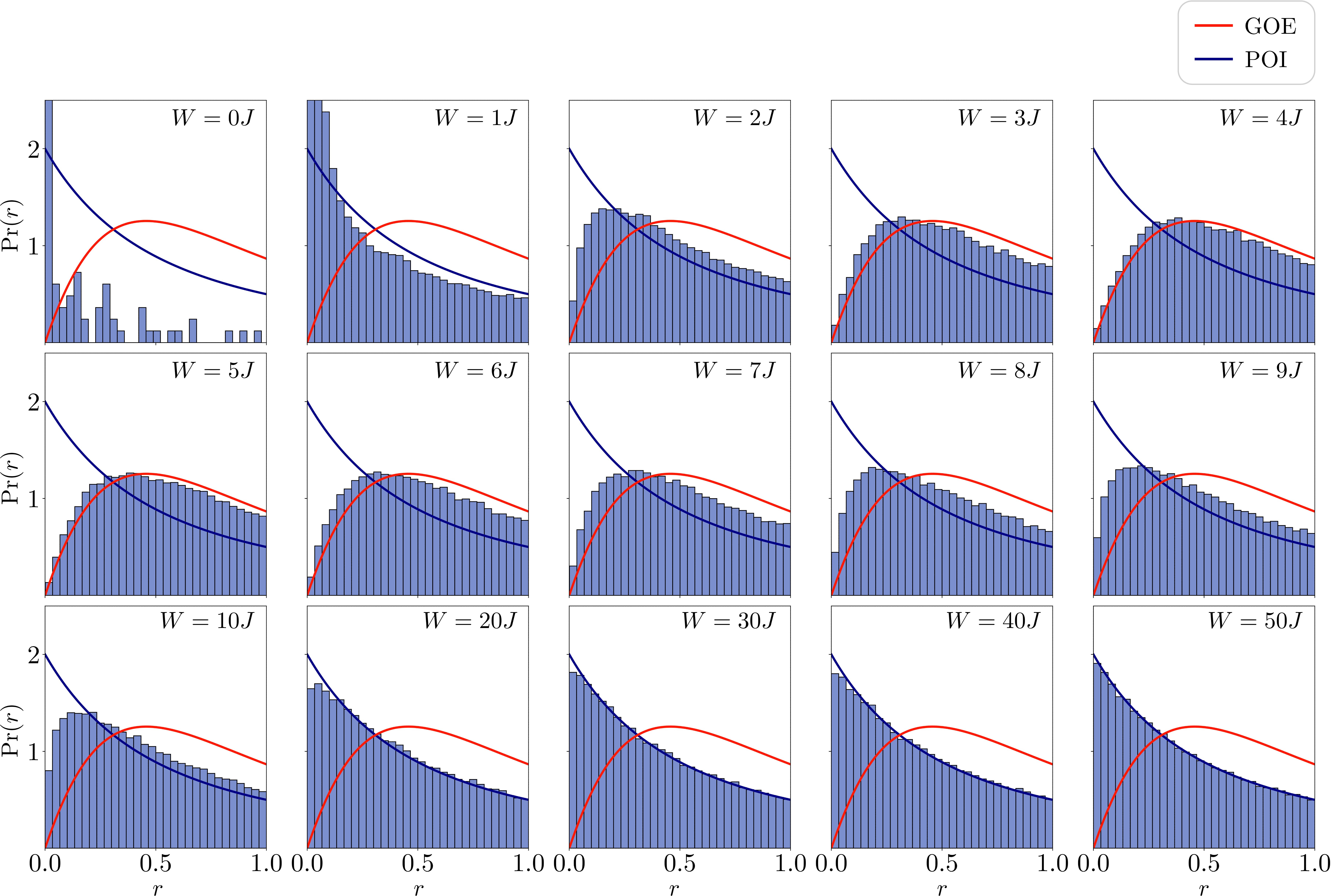}
    \caption{\textbf{Crossover in level statistics from $W=0J$ to $W=50J$.} The histograms represent the level statistics of 500 instances of 9-qubit systems governed by the Hamiltonian in Eq.~\eqref{eq:IsingModel}. At $W=0J$, the system is integrable, leading to the near-degenerate energy levels. As $W$ increases from $3J$ to $50J$, the level statistics evolves from GOE statistics (red curve), indicative of the thermalized phase, to the POI (blue curve), which characterize the MBL phase. The thermalized regime appears around $W=3J$ to $7J$, while the level statistics approach to the Poisson distribution at around $W=20J$. }
    \label{fig:level_stat_vary_W}
\end{figure}

To characterize the dynamical phases in our system governed by Eq.~\eqref{eq:IsingModel}, we compute the level statistics, varying from the disorder strength $W=0J$ to $W=50J$, as shown in Fig.~\ref{fig:level_stat_vary_W}. In the absence of the disorder, $W=0J$, the level statistics deviates from both GOE statistics and POI distribution and the level spacing concentrates around 0, reflecting the highly degenerate energy levels and indicating the phase is neither thermalized nor MBL. Around $W=3J$, the level statistics begin to align with the GOE statistics, showing the emergence of the thermalized phase. As $W$ increases, the level statistics get closer to the GOE statistics and reaches the peak at $W=5J$, before the characteristics of the thermalized phase fade beyond this point. With further increase in the disorder strength, the distribution gradually shifts toward the Poisson distribution. The Poisson distribution is well-approximated for $W > 20J$, suggesting that the system is in the MBL phase in this strong disorder regime. Based on this transition, we select values $W$ where the level statistics are most closely aligned with the GOE statistics and the Poisson distribution to define the thermalized and MBL initialisations, respectively.

\section{Proof of Eq.~\eqref{eq:variance-expressivity-bound}}
\label{appx:proof_bound}
We derive the expressivity-dependent upper bound for the variance of the loss function in Eq.~\eqref{eq:variance-expressivity-bound}. Recall that the parametrized dynamics form an ensemble $\Ubb_{\Theta_D}$, and denote $\rho_0$ as the initial state and $O$ as the observable. The variance of the loss can be expressed as
\begin{align}
    \Var_{\thv\sim \PC}[\LC(\thv)]&=\mathbb{E}_{\thv\sim \PC}[\LC(\thv)^2]-\mathbb{E}_{\thv\sim \PC}[\LC(\thv)]^2 \\
    & = \int_{U \in \Ubb_{\Theta_D}} d\mu(U) \Tr\left[ U^{\otimes2} \rho_0^{\otimes 2} U^{\dagger \otimes 2} O^{\otimes 2}  \right] - \left( \int_{U \in \Ubb_{\Theta_D}} d\mu(U)  \Tr[U\rho_0 U^\dagger O]\right)^2\;, \label{eq:proof-00}
\end{align}
where we use $\Tr[A]\Tr[B] = \Tr[A\otimes B]$ in the second equality.

Next, we consider the operator $\AC_{\Ubb_{\Theta_D}}^{(1)}(\rho_0)$ and $\AC_{\Ubb_{\Theta_D}}^{(2)}(\rho_0)$ of the form
\begin{align}
\AC_{\Ubb_{\Theta_D}}^{(1)}(\rho_0) & = \int_{V\in \,\Ubb(2^n)} d\mu_H(V) V \rho_0 V^{\dagger} - \int_{U \in \Ubb_{\Theta_D}} d\mu(U)  U \rho_0 U^{\dagger} \;, \\
    \AC_{\Ubb_{\Theta_D}}^{(2)}(\rho_0) & = \int_{V\in \,\Ubb(2^n)} d\mu_H(V) V^{\otimes 2} \rho_0^{\otimes 2} V^{\dagger \otimes2} - \int_{U \in \Ubb_{\Theta_D}} d\mu(U)  U^{\otimes2} \rho_0^{\otimes 2} U^{\dagger \otimes 2}  \;,  \label{eq:proof-expressivity-operator} 
\end{align}
where $d\mu_H(V)$ is the Haar measure over the unitary group $\Ubb(2^n)$ of size $2^n$. $\AC_{\Ubb_{\Theta_D}}^{(2)}(\rho_0)$ and $\AC_{\Ubb_{\Theta_D}}^{(2)}(\rho_0)$ indicate how far away the first and second moment operators on $\rho_0$ over $\Ubb_{\Theta_D}$ are from the Haar random.

By expressing Eq.~\eqref{eq:proof-00} in terms of the operators $\AC_{\Ubb_{\Theta_D}}^{(1)}(\rho_0)$ and  $\AC_{\Ubb_{\Theta_D}}^{(2)}(\rho_0)$, we have
\begin{align}
    \Var_{\thv\sim \PC}[\LC(\thv)]=& \int_{V\in \,\Ubb(2^n)} d\mu_H(V) \Tr\left[ V^{\otimes 2} \rho_0^{\otimes 2} V^{\dagger \otimes2} O^{\otimes 2}\right] -  \Tr\left[ \AC_{\Ubb_{\Theta_D}}^{(2)}(\rho_0) O^{\otimes 2}\right] \nonumber \\
    & - \left( \int_{V\in \,\Ubb(2^n)} d\mu_H(V) \Tr\left[ V \rho_0 V^{\dagger } O\right] -  \Tr\left[ \AC_{\Ubb_{\Theta_D}}^{(1)}(\rho_0) O\right]\right)^2 \\
    = & \Var_{\rm Haar}[\LC(\thv)] - \Tr\left[ \AC_{\Ubb_{\Theta_D}}^{(2)}(\rho_0) O^{\otimes 2}\right] + \frac{\Tr[O]}{2^{n-1}} \Tr\left[ \AC_{\Ubb_{\Theta_D}}^{(1)}(\rho_0) O\right] - \left(\Tr\left[ \AC_{\Ubb_{\Theta_D}}^{(1)}(\rho_0) O\right] \right)^2 \label{eq:proof-step-var}\\
    \leq & \Var_{\rm Haar}[\LC(\thv)]  + \| \AC_{\Ubb_{\Theta_D}}^{(2)} (\rho_0)\|_2 \| O\|^2_2 + \frac{\Tr[O]}{2^{n-1}} \| \AC_{\Ubb_{\Theta_D}}^{(1)}(\rho_0) \|_2 \| O\|_2 - \| \AC_{\Ubb_{\Theta_D}}^{(1)}(\rho_0) \|_2^2 \| O\|_2^2 \;\;,
\label{eq:proof-01}
\end{align}
where to reach the last inequality we take the absolute of the Eq.~\eqref{eq:proof-step-var} and use the triangle inequality $|a+b| < |a| + |b|$ (for some values $a,b$) followed by applying Cauchy-Schwarz inequality $\Tr|AB| \leq \|A\|_2\|B\|_2$ (for some matrices $A,B$) and $\|O^{\otimes2}\|_2 = \|O\|_2^2$.

Next, we link the norms $\| \AC_{\Ubb_{\Theta_D}}^{(1)} (\rho_0)\|_2$ and $\| \AC_{\Ubb_{\Theta_D}}^{(2)} (\rho_0)\|_2$ in Eq.~\eqref{eq:proof-01} to their respective frame potentials. In particular, consider the $t$-order frame potentials with respect to $\mathbb{U}_{\Theta_{D}}$ and Haar random
\begin{equation}
\FC^{(t)}_{\mathbb{U}_{\Theta_{D}}} = \int\int_{U,V\in \mathbb{U}_{\Theta_{D}}} d\mu(U)d\mu(V) |\bra{0}V^\dagger U\ket{0}|^{2t}\;\;,
\end{equation}  and 
\begin{equation}\FC^{(t)}_{\rm Haar} = \int\int_{U,V\in \Ubb(2^n)} d\mu_H(U) d\mu_H(V) |\bra{0}V^\dagger U\ket{0}|^{2t} \;\;.
\end{equation}

Now, let us unroll $\| \AC_{\Ubb_{\Theta_D}}^{(t)} (\rho_0)\|_2$ as follows
\begin{align}
    \| \AC_{\Ubb_{\Theta_D}}^{(t)}(\rho_0) \|^2_2 = & \Tr\Bigg[\left(\int_{V'\in \,\Ubb(2^n)} d\mu_H(V') V'^{\otimes t} \rho_0^{\otimes t} V'^{\dagger \otimes t} - \int_{U' \in \Ubb_{\Theta_D}} d\mu(U')  U'^{\otimes t} \rho_0^{\otimes t} U'^{\dagger \otimes t} \right)^\dagger     \nonumber\\
    & \;\;\;\;\;\;\;\;\left(\int_{V\in \,\Ubb(2^n)} d\mu_H(V) V^{\otimes t} \rho_0^{\otimes t} V^{\dagger \otimes t} 
    - \int_{U \in \Ubb_{\Theta_D}} d\mu(U)  U^{\otimes t} \rho_0^{\otimes t} U^{\dagger \otimes t} \right)\Bigg] \\ 
    =&\int_{V,V'\in \,\Ubb(2^n)} d\mu_H(V) d\mu_H(V') \Tr[(V' \rho_0  V'^{\dagger} V \rho_0 V^{\dagger})^ {\otimes t}]
    + \int_{U,U' \in \Ubb_{\Theta_D}} d\mu(U) d\mu(U') \Tr[(U'\rho_0  U'^{\dagger} U \rho_0 U^{\dagger})^{ \otimes t}] \nonumber
    \\&\;- \int_{V'\in \,\Ubb(2^n)} \int_{U \in \Ubb_{\Theta_D}} d\mu_H(V') d\mu(U) \Tr[(V' \rho_0 V'^{\dagger} U \rho_0 U^{\dagger})^{\otimes t}] \nonumber
    \\&\;-  \int_{U' \in \Ubb_{\Theta_D}} \int_{V\in \,\Ubb(2^n)} d\mu(U') d\mu_H(V) \Tr[ (U' \rho_0  U'^{\dagger } V \rho_0 V^{\dagger}]^{\otimes t})]  \\
    =&\int_{U,U' \in \Ubb_{\Theta_D}} d\mu(U) d\mu(U') |\bra{0} U'^\dagger U \ket{0}|^{2t} - \int_{V,V'\in \,\Ubb(2^n)} d\mu_H(V) d\mu_H(V')|\bra{0} V'^\dagger V \ket{0}|^{2t}\;\;,\\
    =& \FC^{(t)}_{\mathbb{U}_{\Theta_{D}}} - \FC^{(t)}_{\rm Haar}\;\;, \label{eq:proof-frame-norm}
\end{align}
where in the second equality we expand the product and repeatedly use $(A\otimes B)(C\otimes D) = AC \otimes BD$ (for matrices $A,B,C,D$). To reach the third equality we repeatedly use the invariant of the Haar measure, i.e., $\int_{V\in \,\Ubb(2^n)}d\mu_H(V)f(V) = \int_{V\in \,\Ubb(2^n)}d\mu_H(V) f(UV) =  \int_{V\in \,\Ubb(2^n)}d\mu_H(V) f(VU)$
for any arbitrary unitary $U$ and $\rho_0 = |0\rangle\langle0|$. In particular, we have
\begin{align}
\int_{V\in \,\Ubb(2^n)}d\mu_H(V)\Tr[VU\rho_0 U^\dagger V^\dagger] &=\int_{V\in \,\Ubb(2^n)}d\mu_H(V)\Tr[(VU)\rho_0 (V U)^\dagger] \;\;,\\
&= \int_{V\in \,\Ubb(2^n)}d\mu_H(V)\Tr[V\rho_0 V^\dagger]\;\;,
\end{align}
while in the fourth inequality, we use $\rho_0 =|\vec{0}\rangle\langle \vec{0}|$. 

Thus, by plugging Eq.~\eqref{eq:proof-frame-norm} for $t=1,2$ to Eq.~\eqref{eq:proof-01}, we have 
\begin{equation}
\Var_{\thv\sim \PC}[\LC(\thv)]\leq \Var_{\rm Haar}[\LC(\thv)]  + \left( \sqrt{\FC^{(2)}_{\mathbb{U}_{\Theta_{D}}} - \FC^{(2)}_{\rm Haar}}-\FC^{(1)}_{\mathbb{U}_{\Theta_{D}}} + \FC^{(1)}_{\rm Haar}\right) \| O\|^2_2 + \frac{\Tr[O]}{2^{n-1}} \sqrt{\FC^{(1)}_{\mathbb{U}_{\Theta_{D}}} - \FC^{(1)}_{\rm Haar}} \| O\|_2 \;,
\end{equation}
which is the promised bound in Eq.~\eqref{eq:variance-expressivity-bound} and completes the proof.

\section{Disordered Ising Hamiltonian with long-range interactions}\label{appx:MoreRealisticModel}
\begin{figure}[h!]
    \centering
    \includegraphics[width=0.7\linewidth]{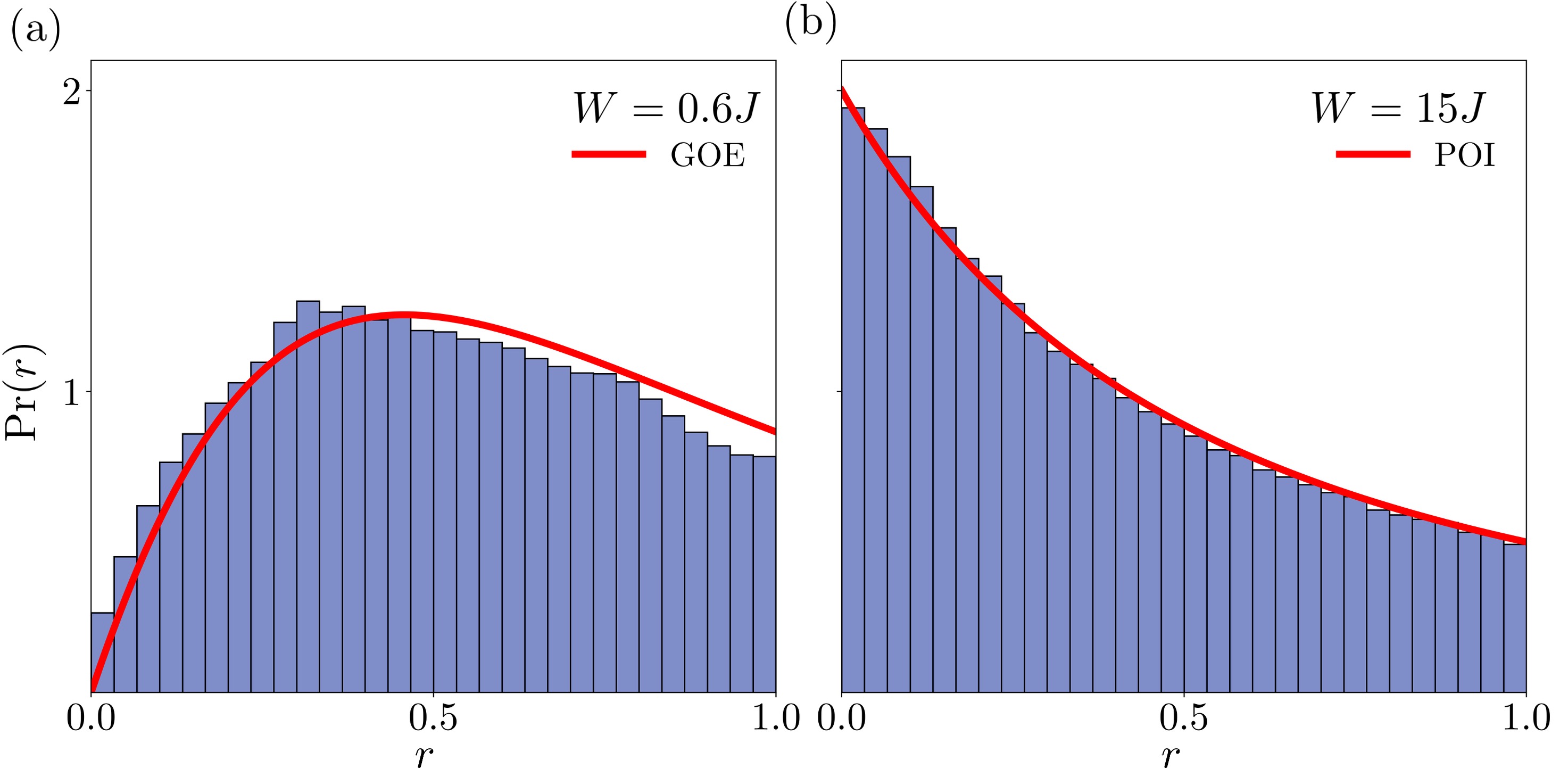}
    \caption{\textbf{Level statistics of the disordered Ising model with long-range interactions.} The histograms represent the level statistics of 500 instances of 9-qubit systems governed by the Hamiltonian in Eq.~\eqref{eq:expising} with two disorder strengths. (Left) With $W=0.6J$, the histogram follows the GOE statistics, indicative of the system being in the thermalized phase. (Right) In contrast, with $W=15J$, the histogram follows the Poisson distribution, indicative of the system being in the MBL phase. These different disorder strengths are used for initialising the parameters in our ans\"{a}tze for the thermalized and MBL initialisation.}
    \label{fig:exp_level_stat}
\end{figure}

\begin{figure}
    \centering
    \includegraphics[width=\linewidth]{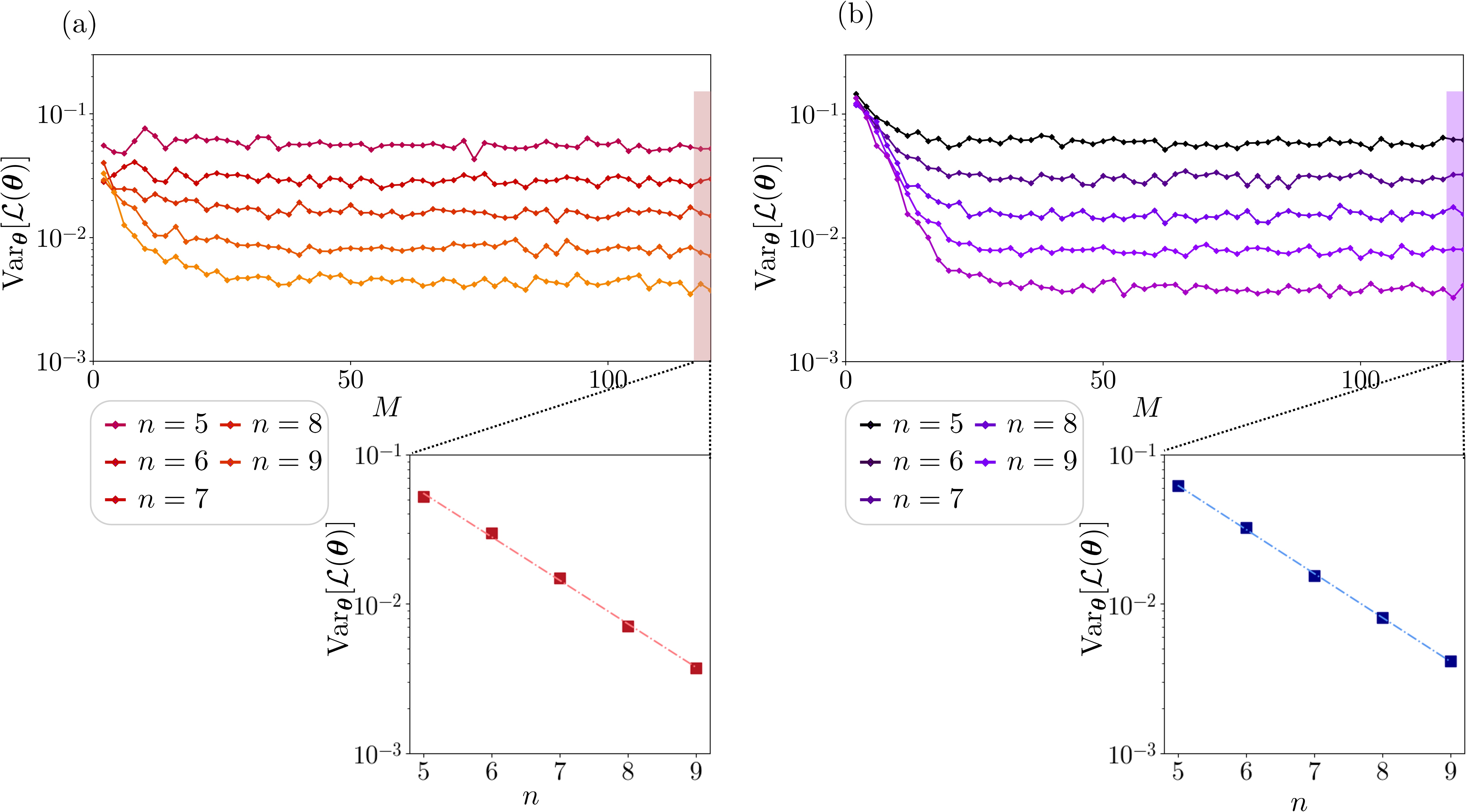}
    \caption{\textbf{The onset of barren plateaus 
    of the thermalized and MBL initialisations.} (a) The variance of $\expval{Z_1Z_2}_{\vec{\theta}}$ is plotted against the number of quenches for systems with 5 to 9 qubits governed by the Hamiltonian in Eq.~\eqref{eq:expising}, averaged over 400 realizations, for (Left) the thermalized and (Right) the MBL initialisations. (c) and (d) panels show the saturated variance values of $\langle Z_1Z_2 \rangle_{\vec{\theta}}$ for both phases, confirming the emergence of barren plateaus for the thermalized and MBL initialisation, respectively. }
    \label{fig:BP_compare_experiment}
\end{figure}

We investigate a more realistic model inspired by experimental trapped-ion systems, as described in Ref.~\cite{smith2016many}. Specifically, we explore the system evolving under the dynamics in Eq.~\eqref{eq:quench_dynamics}, governed by a disordered Ising model with a long-range interaction Hamiltonian, and interchange the two Pauli operators, $X \leftrightarrow{Z}$, in the original Hamiltonian:
\begin{equation}
    \label{eq:expising}
        H(\thv) = J\sum_{i>j}\frac{Z_iZ_j}{|i-j|^\alpha}+B\sum_iX_i+ \sum_i h_iX_i,
\end{equation} 
where $J$ is the characteristic interaction strength, $\alpha$ controls the strength of the long-range interaction, $B$ represents a uniform effective transverse field, and $\thv = \{h_i\}^n_{i=1}$ are the on-site disordered energies. This system can exhibit both thermalized and MBL phases. Using the same procedure as in section~\ref{subsec:phase}, we identify the thermalized phase for $W=0.6J$ and the MBL phase for $W=15J$, given $\alpha = 1$, $B=6J$, and $h_i$ uniformly sampled from the interval $[-W/2,W/2]$. These phases are evident from the level statistics shown in Fig.~\ref{fig:exp_level_stat}. 
We investigate barren plateaus (BPs) as in the main text. However, to align with experimental setups, we initialise the quantum state in the Néel ordered state along the $x$-direction, represented as $|+-+-\cdots\rangle$ in the computational basis, where $|+\rangle = \frac{|0\rangle+|1\rangle}{\sqrt{2}}$ and $|-\rangle=\frac{|0\rangle-|1\rangle}{\sqrt{2}}$. As a result, we observe that the variance of $\langle Z_1 Z_2 \rangle$ decays as functions of the number of quenches, as shown in Fig.~\ref{fig:BP_compare_experiment}. While the variance in the thermalized initialisation decays relatively faster than in the MBL initialisation, the difference between the onset of BPs in the two initialisations is much smaller compared to the model in the main text. This suggests that although there is an advantage in the MBL initialisation over the thermalized one in this model, the smaller range of the advantageous regime in this more realistic model raises questions about the influence of the long-range interaction and the choice of the initial state.

\section{Numerical simulations to demonstrate MBL initialisation capability for solving optimization tasks}
\label{appx:extended_numerical_result}

To demonstrate the capability of the proposed MBL initialisation scheme, we use two different ansätzes, governed by the experimentally-feasible Hamiltonian in Eq.~\eqref{eq:expising} and the Ising Hamiltonian in Eq.~\eqref{eq:IsingModel},  to solve two optimization tasks: VQE and Max-Cut problems.  Furthermore, at the end of this appendix, we investigate the optimization results across different problem sizes, with the optimization settings remaining consistent with those detailed in Section~\ref{sec:demonstration}.

\subsection{Utility of MBL initialisation in different models}
\paragraph{VQE:}
Here, we employ 100 instances of our ans\"{a}tze--a parametrized quench dynamics initialised in the MBL phase--to solve VQE problems in two examples. We select the number of quenches in the ans\"{a}tze to be either 2 or 6, as these different numbers of quenches lead to noticeable differences in expressivity and the variance of the loss function.

\medskip
\noindent\textbf{Example 1:} We use an ans\"{a}tze governed by the Hamiltonian in Eq.~\eqref{eq:IsingModel} to find the ground state of a 7-qubit long-range Ising model Hamiltonian of the form in Eq.~\eqref{eq:expising} within the thermalized phase. After 100 training iterations: 
\begin{itemize}
    \item \textbf{2-quench ans\"{a}tze:} Average relative error is $0.68\%$ with a standard deviation of $1.1\%$.
    \item \textbf{6-quench ans\"{a}tze:} Average relative error is $0.48\%$ with a standard deviation of $0.57\%$.
\end{itemize}

\noindent\textbf{Example 2:} We use the ans\"{a}tze from the Hamiltonian in Eq.~\eqref{eq:expising} to solve a 7-qubit nearest-neighbour Ising Hamiltonian of the form in Eq.~\eqref{eq:IsingModel} within the thermalized phase. After 100 training iterations:
\begin{itemize}
    \item \textbf{2-quench ans\"{a}tze:} Average relative error is $3.1\%$ with a standard deviation of $2.6\%$.
    \item \textbf{6-quench ans\"{a}tze:} Average relative error is $0.89\%$ with a standard deviation of $0.51\%$.
\end{itemize}

\begin{figure}[h!]
        \centering
        \includegraphics[width=0.8\linewidth]{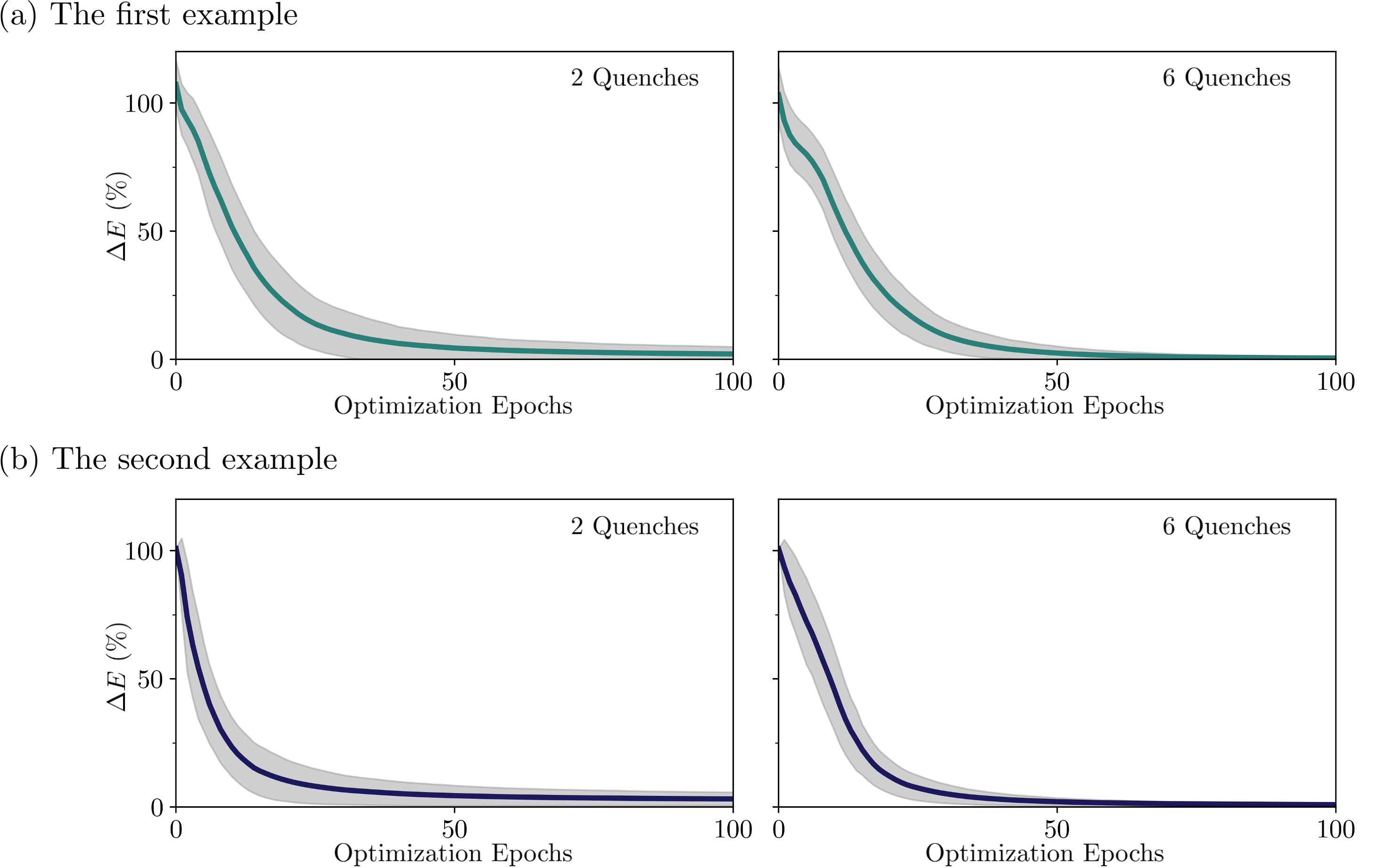}
        \caption{\textbf{VQE optimization results using different ansätzes.} 
        The average relative error between the estimated ground state energy, $\expval{H}$, and the actual ground state energy of the Hamiltonian in (a) Eq.~\eqref{eq:expising} and (b) Eq.~\eqref{eq:IsingModel} are plotted on a logarithmic scale against training iteration. The plots represent the performance of the ans\"{a}tze given by the Hamiltonian in (a) Eq.~\eqref{eq:IsingModel} and (b) Eq.~\eqref{eq:expising}, initialised in the MBL phase over 100 realizations, using (Left) 2 quenches and (Right) 6 quenches. The shaded grey areas indicate the standard deviation for each training iteration.
        }
        \label{fig:VQE}
\end{figure}

The optimization processes for both examples are shown in Fig.~\ref{fig:VQE}. In both cases, increasing the number of quenches from 2 to 6 improves the accuracy of the estimated ground state energy. This suggests that increasing the number of quenches can enhance optimization performance to some extent. 

\medskip

\paragraph{Max-Cut Problem:}
For demonstration, we consider a specific graph instance associated with the adjacency matrix:
\begin{equation}
    \begin{bmatrix}
        0 & 1 & -1 & 0 & 1\\
        1 & 0 &  1 & 0 & 0\\
        -1 & 1 &  0 & -1 & 1\\
        0 & 0 &  -1 & 0 & -1\\
        1 & 0 &  1 & -1 & 0
    \end{bmatrix}.
\end{equation}
This graph has multiple optimal solutions, corresponding to the four-fold degenerate ground state configurations: $\ket{01001}$, $\ket{10110}$, $\ket{10010}$, and $\ket{10110}$.

\begin{figure}
        \centering
        \includegraphics[width=\linewidth]{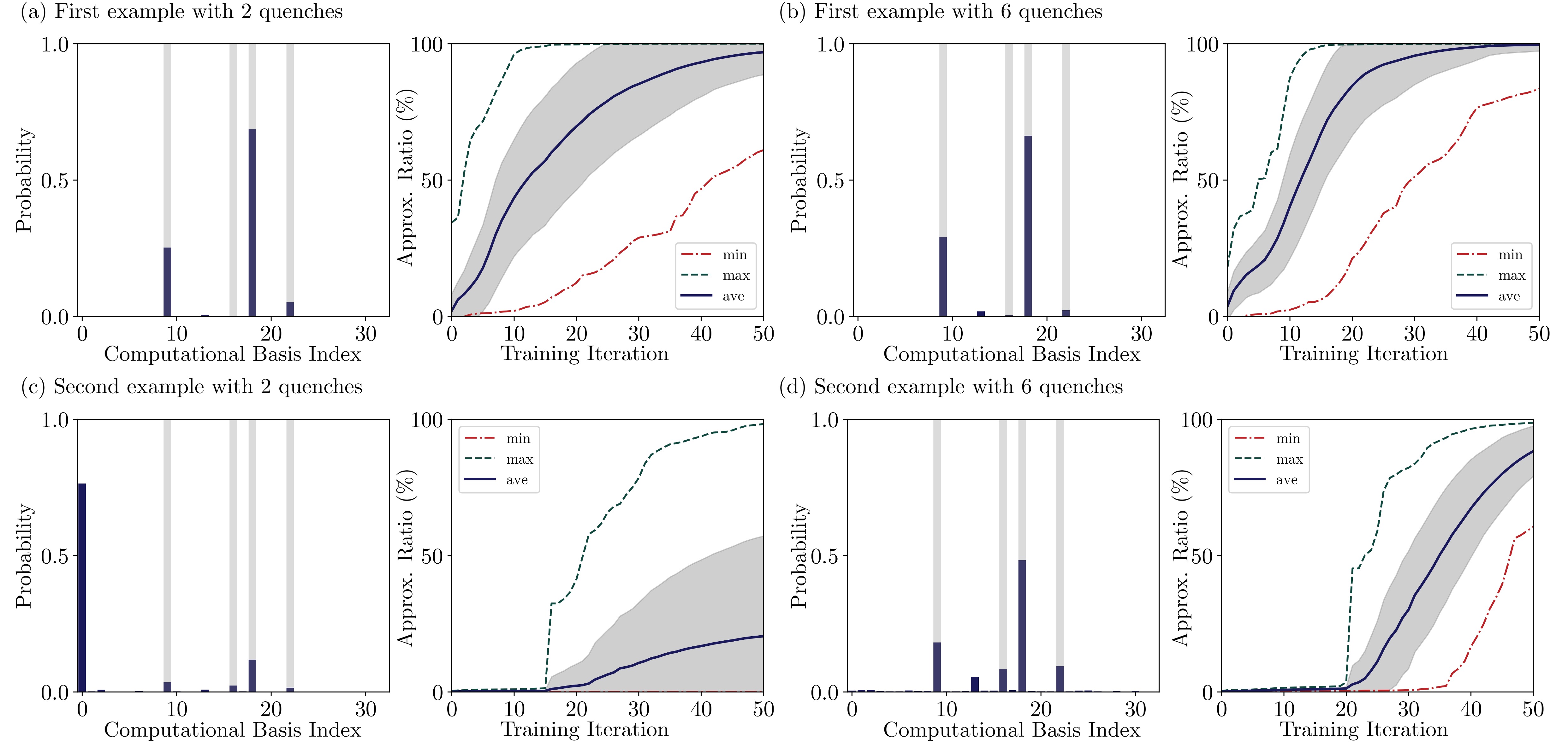}
        \caption{\textbf{Max-Cut optimization results using different ansätzes.} 
        The probability distribution of each computational basis in the output state from the quench dynamics governed by the Hamiltonian in (a) Eq.~\eqref{eq:IsingModel} and (b) Eq.~\eqref{eq:expising} after optimization over 50 training iterations, and the average approximation ratio plotted against training iteration, averaged over 100 realizations, are shown for (Left) the 2-quench case and (Right) the 6-quench case. The four vertically shaded gray columns highlight the four-fold degenerate ground states of this Max-Cut instance. The shaded gray regions surrounding the average approximation ratios indicate the standard deviations of these ratios.}\label{fig:QUBO_optimization}
\end{figure}

Here, we employ 100 instances of our MBL-initialised ans\"{a}tzes to solve the problem. We consider two cases: 2-quench and 6-quench ans\"{a}tzes for the same reason as in the previous subsection, and choose the selection probability threshold of $0.01$. The results after 50 training iterations are shown in Fig~\ref{fig:QUBO_optimization}.

\medskip
\noindent\textbf{Example 1: the nearest-neighbour Ising Hamiltonian ans\"{a}tze in Eq.~\eqref{eq:IsingModel}} 
\begin{itemize}
    \item \textbf{2-quench ans\"{a}tze:} Average approximation ratio is $99.59\%$ with a standard deviation of $2.30\%$.
     \item \textbf{6-quench ans\"{a}tze:} Average approximation ratio is $99.61\%$ with a standard deviation of $2.27\%$.
\end{itemize}
Both ans\"{a}tzes successfully identify the degenerate ground states.

\medskip
\noindent\textbf{Example 2: the long-range Ising Hamiltonian ans\"{a}tze in Eq.~\eqref{eq:expising}}
\begin{itemize}
    \item \textbf{2-quench ans\"{a}tze:} Average approximation ratio is $20.45\%$ with a standard deviation of $36.73\%$.
     \item \textbf{6-quench ans\"{a}tze:} Average approximation ratio is $88.34\%$ with a standard deviation of $9.23\%$.
\end{itemize}
While increasing the number of quenches in the ans\"{a}tze shows improvement, the performance still falls short compared to Example 1. 

\subsection{Extended numerical results of initialisation strategies across different system sizes}
With the same setup as in  Section~\ref{sec:demonstration}, we test the optimization results on different problem sizes. In particular, we use shallow quenches $M=2$ for the thermalized initialisation, and intermediate quenches $M=22$ (for $n=6$), $26$ (for $n=7$), $24$ (for $n=8$), and $26$ (for $n=10$), for the MBL initialisation.

\begin{figure}[h!]
        \centering
        \includegraphics[width=\linewidth]{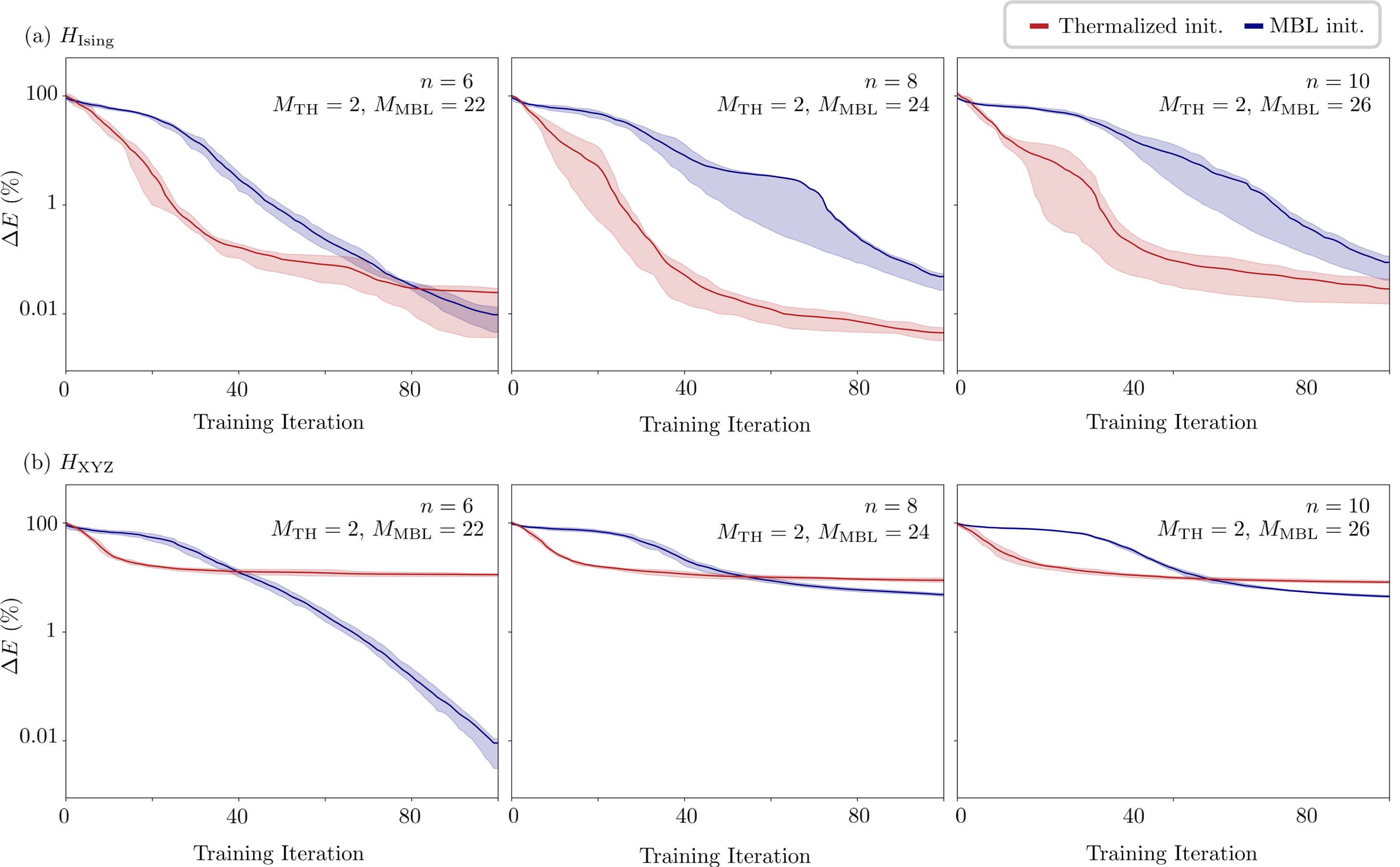}
        \caption{\textbf{Extended VQE optimization results.}
         The average relative error over $5$ initialisations is plotted against the training iteration on a logarithmic scale for system sizes $n=6$, $n=8$ and $n=10$. The target Hamiltonians are instances of the long-range disordered Ising model and the Heisenberg model with periodic boundary conditions. The red line represents the optimization process using thermalized initialisation with $M=2$ and the blue line shows the optimization process using the MBL initialisation with $M=22$ for $n=6$, $M=24$ for $n=8$, and $M=26$ for $n=10$. The shaded region around each line represents the range between $20^\text{th}$ and $80^\text{th}$ percentiles of the relative error.}
        \label{fig:extended_VQE_result}
\end{figure}

\begin{table}[h]
    \centering
    \begin{tabular}{|c|c|cc|cc|cc|}
        \hline
        \multirow{2}{*}{\textbf{Hamiltonian}} & \multirow{2}{*}{\textbf{Initialisation}} & \multicolumn{2}{c|}{\boldmath$n=6$} & \multicolumn{2}{c|}{\boldmath$n=8$} & \multicolumn{2}{c|}{\boldmath$n=10$} \\
        \cline{3-8}
        && \textbf{Mean (\%)} & \textbf{Std. (\%)} & \textbf{Mean (\%)} & \textbf{Std. (\%)}  & \textbf{Mean (\%)} & \textbf{Std. (\%)} \\
        \hline
        \multirow{2}{*}{\boldmath$H_{\text{Ising}}$} & MBL & $9.65\times10^{-3}$ & $6.72\times{10^{-3}}$ & $4.83\times10^{-2}$ & $3.28\times10^{-2}$ & $8.88\times10^{-2}$ & $4.70\times10^{-2}$  \\
        \cline{2-8}
        & TH & $2.47\times10^{-2}$ & $3.72\times10^{-2}$ &  $4.51\times10^{-3}$ & $1.60\times10^{-3}$ &  $2.87\times10^{-2}$ & $1.83\times10^{-2}$ \\
        \hline
        \multirow{2}{*}{\boldmath$H_{\text{XYZ}}$} & MBL & $9.15\times10^{-3}$ & $1.10\times10^{-2}$ & $4.82$ & $0.401$  &$4.47$ & $0.226$ \\
        \cline{2-8}
        & TH &$11.2$& $1.34$ &  $8.84$ & $0.785$ & $8.15$ & $0.869$ \\
        \hline
    \end{tabular}
    \caption{VQE optimization results for $H_\text{Ising}$ and $H_\text{XYZ}$ across different system sizes ($n=6$, $n=8$, and $n=10$) averaged over $5$ initialisation instances after $100$ training iterations. The performance is measured by the average and the standard deviation of the relative errors.}
    \label{tab:VQE_result}
\end{table}
We have shown the result of optimization for two different target Hamiltonians over $100$ training iterations in Table~\ref{tab:VQE_result} and Fig.~\ref{fig:extended_VQE_result}. For $H_\text{Ising}$, the thermalized initialisation outperforms the MBL initialisation in $n=8$ and $n=10$ systems. For $H_\text{XYZ}$, the MBL consistently achieves better performance across all system sizes. However, we observe that for $n=8$ and $n=10$ with only $100$ training iterations, the improvements are not significant. It is important to note that the accuracy of these optimizations is significantly improved for the case $n=8$ as previously demonstrated in the main text.

\section{Choice of different time evolution and its effect on the variance scaling}
\label{appx:diff_time}
\begin{figure}[h!]
        \centering
        \includegraphics[width=\linewidth]{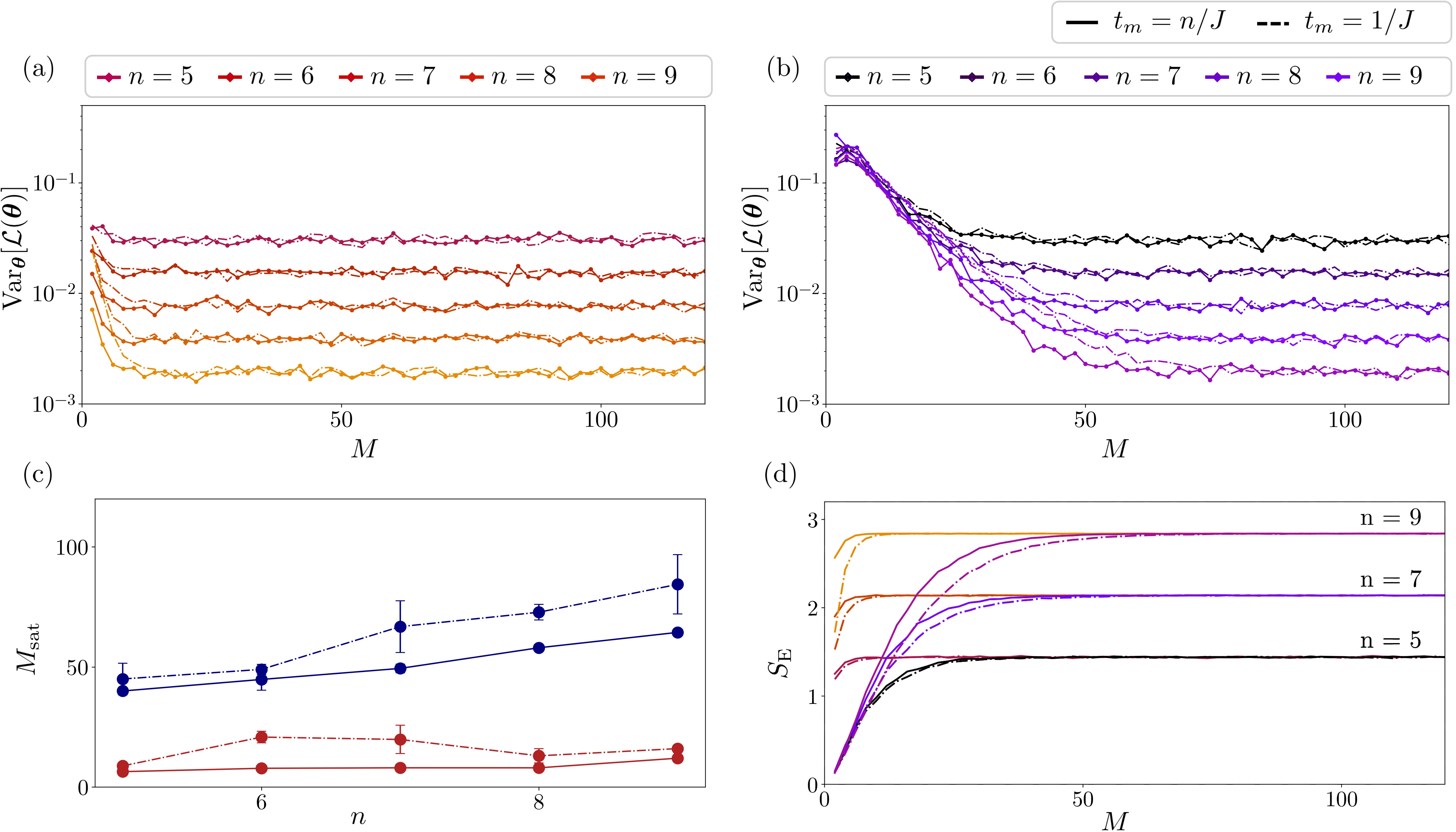}
        \caption{\textbf{Effect of time scale on BPs and entanglement entropy.} Two different evolution times within each quench are considered: $t_m=n/J$ (solid lines) and $t_m=1/J$ (dashed lines). The top panels show the variance of the loss function for (a) thermalized and (b) MBL initialisations as the number of quenches increases. Panel (c) shows the number of quenches required for the onset of BP ($M_\text{sat}$) for both thermalized (red) and MBL (blue) initialisations. The bottom-right panel displays the bipartite entanglement entropy versus the number of quenches. Data is for systems with 5 to 9 qubits.
        }
        \label{fig:BP_time}
\end{figure}

We conduct an investigation on the effect of the evolution time in each quench, $t_m$. By setting $t_m=n/J$, we study how correlation spreading within a single quench can accelerate the variance convergence. Indeed, we observe that the entanglement grows faster than in the case of the original time scale, $t_m=1/J$, as shown in Fig.~\ref{fig:BP_time}~(d). Consequently, as discussed in Section~\ref{sec:BP}, a smaller number of quenches is required for the cost landscape to enter the BP regime. This effectively squeezes the regime II, as  studied in Section~\ref{sec:strategy}. This expectation is confirmed in Fig.~\ref{fig:BP_time}~(a)-(c), comparing the number of quenches at which BP onset occurs for different $t_m$. Nevertheless, while the number of quenches is reduced, this does not qualitatively change the fundamental findings or the overall concept of the initialisation strategy.

\end{document}